\documentclass[aps,twocolumn,prb,10pt,superscriptaddress]{revtex4-1}

\usepackage{xcolor}
\usepackage{amsmath}
\usepackage{graphicx}
\usepackage{grffile}
\usepackage{xspace}
\usepackage{ulem}

\graphicspath{{figs/}}

\newcommand{\lanio}{LaNiO$_{3}$\xspace}

\newcommand{\eps}{\varepsilon}

\newcommand{\iomn}{i\omega_{n}}
\newcommand{\kv}{\mathbf{k}}

\newcommand{\cf}{\Delta_{\mathrm{CF}}}

\newcommand{\pol}{P}
\newcommand{\psusc}{\chi_{P}}
\newcommand{\Pav}{P_{\mathrm{av}}}

\newcommand{\xxyy}{x^{2} - y^{2}}
\newcommand{\zz}{z^{2}}
\newcommand{\eg}{e_{g}}
\newcommand{\ttg}{t_{2g}}
\newcommand{\dlbar}{d^{8}\underline{L}}

\newcommand{\bw}{\textrm{LW}}
\newcommand{\sw}{\textrm{EW}}

\newcommand{\ket}[1]{\left.\left|{#1}\right\rangle\right.}

\newcommand{\GlocSW}{G^{\mathrm{loc,EW}}}
\newcommand{\GlocBW}{G^{\mathrm{loc,LW}}}
\newcommand{\Gband}{G^{\mathrm{band}}}
\newcommand{\Sigloc}{\Sigma^{\mathrm{loc}}}

\def\lno{LaNiO$_3$\xspace}
\newcommand{\tetra}{tetragonal\xspace}
\newcommand{\dist}{distorted\xspace}

\begin{document}

\title{Orbital polarization in strained \lno: Structural distortions and correlation effects}

\author{Oleg E. Peil}\email{Oleg.Peil@gmail.com}
\affiliation{D\'epartement de Physique de la Mati\`ere Condens\'ee, University of Geneva,
24 Quai Ernest-Ansermet, 1211 Gen\`eve 4, Switzerland}
\affiliation{Centre de Physique Th\'eorique, Ecole Polytechnique, CNRS, 91128 Palaiseau Cedex, France}
\author{Michel Ferrero}
\affiliation{Centre de Physique Th\'eorique, Ecole Polytechnique, CNRS, 91128 Palaiseau Cedex, France}
\author{Antoine Georges}
\affiliation{D\'epartement de Physique de la Mati\`ere Condens\'ee, University of Geneva,
24 Quai Ernest-Ansermet, 1211 Gen\`eve 4, Switzerland}
\affiliation{Centre de Physique Th\'eorique, Ecole Polytechnique, CNRS, 91128 Palaiseau Cedex, France}
\affiliation{Coll\`ege de France, 11 place Marcelin Berthelot, 75005 Paris, France}

\begin{abstract}
%
%
Transition-metal heterostructures offer the fascinating possibility of controlling orbital
degrees of freedom via strain. Here, we investigate theoretically the degree of
orbital polarization that can be induced by epitaxial strain in LaNiO$_3$ films.
Using combined electronic structure and dynamical mean-field theory methods
we take into account both structural distortions and electron correlations and discuss
their relative influence. We confirm that Hund's rule coupling tends to decrease
the polarization and point out that this applies to both the $\dlbar$ and $d^7$
local configurations of the Ni ions. Our calculations are in good agreement with
recent experiments, which revealed sizable orbital polarization under tensile strain.
We discuss why full orbital polarization is hard to achieve in this specific
system and emphasize the general limitations that must be overcome to achieve this goal.
\end{abstract}

\maketitle

%
%
\section{Introduction}
\label{sec:intro}

Ultrathin films and heterostructures of transition-metal oxides (TMOs) have attracted considerable interest in the past decade. 
Recent advances in TMO heterostructure and strain engineering have provided 
increased control of the electronic properties of TMOs. Furthermore, these structures exhibit novel behavior 
not found in their bulk counterparts.\cite{Schlom2007,Zubko2011,Rondinelli2011,Hwang2012,Gibert2012,Boris2011}

The family of rare-earth nickelates,\cite{Medarde1997,Catalan2008} RNiO$_3$, has attracted particular 
attention in this context. Indeed, this class of materials has a rich phase diagram displaying a
metal-to-paramagnetic-insulator, as well as a metal-to-magnetic-insulator transitions. 
These transitions are affected by the structural distortions depending, in turn, on 
the radius of the rare-earth ion.
This interplay between structural and electronic properties makes this class of materials 
particularly suitable for heterostructure and strain engineering.

In this paper, we focus on \lanio (LNO). In bulk equilibrium form, this material is an 
exception among the RNiO$_3$ family since it remains a paramagnetic metal down to
the lowest temperatures.\cite{Sreedhar1992, Xu1993}
This metal has a rather high degree of electronic correlations, however, as 
signaled by the enhancement of the effective mass and susceptibility as compared 
to band values, as well as the sizable $T^2$ coefficient of the resistivity.
\cite{Rajeev1991, Xu1993, Eguchi2009, Ouellette2010}
One may thus expect that this material can be rather easily tuned to become an insulator.
Indeed, ultrathin LNO films were shown to become insulating  
under either dimensionality control \cite{Boris2011,Gray2011,Scherwitzl2009} or
epitaxial strain.\cite{Son2010,Scherwitzl2011}
It was also demonstrated that both strain and dimensional confinement can drive
an LNO film towards a spin-density-wave state, which is similar to
what is observed in the insulating phase of other nickelates.\cite{Frano2013}
This makes \lanio a very suitable system for materials design by strain
engineering and heterostructuring.   

In a pioneering article, Chaloupka and Khaliullin\cite{Chaloupka2008} proposed that strained 
heterostructures of LNO could be used to engineer a material having
an electronic structure consisting of a single band crossing the Fermi level. 
In view of the strong electronic correlations in the Ni $d$-shell, the low-energy effective model describing 
such a material might thus be quite analogous to the one appropriate for cuprates, hence suggesting a 
favorable situation for strong superexchange and possible high-$T_c$ superconductivity. 
Indeed, the low-energy electronic structure at the Fermi level of bulk \lanio is primarily
determined by the Ni$^{3+}$ degenerate $\eg$ states which form a two-sheet 
Fermi surface.\cite{Eguchi2009,Deng2012}
One of the major effects of epitaxial strain or heterostructuring
is the degeneracy lifting of the $\eg$ states, resulting in 
`orbital polarization' (OP) of the electronic structure. 
The key question is whether conditions can be found such that this OP is large and 
the quasi two-dimensional $d_{\xxyy}$ band is a dominantly occupied one with proper filling.  
This issue was previously investigated by Hansmann \textit{et al.}\cite{Hansmann2009, Hansmann2010} 
and Han \textit{et al.},\cite{Han2011} for LaNiO$_3$/LaAlO$_3$ 
heterostructures, and proposals for achieving high OP by using other 
LNO-based heterostructures\cite{Chen2013,Chen2013b,Ruegg2013,Doennig2014} or chemical control 
by other counterions\cite{Yang2010, Han2010} were made. 

From a theoretical standpoint, the degree of orbital polarization has been a subject of controversy, 
mainly due to the need for a proper treatment of both strong correlation effects 
in the Ni $d$-shell and
of the strong hybridization 
effects with oxygen ligands.\cite{Anisimov1999,Mizokawa2000}
In the work of Hansmann \textit{et al.},\cite{Hansmann2009, Hansmann2010} a low-energy 
description involving only the two $e_g$-like bands occupied by a single electron
was considered
(corresponding to the nominal $d^7$ occupancy of the Ni $d$-shell). 
It was concluded there that correlation effects may lead to a considerable 
enhancement of the OP, mostly due to the effect of the on-site Coulomb repulsion 
$U$, as emphasized in previous model studies.\cite{Poteryaev2008}
This conclusion was challenged by Han \textit{et al.},\cite{Han2011} who pointed out that the joint effect of the 
Hund's rule coupling and of the strong hybridization with the ligand (associated with the relevance 
of the $\dlbar$ configuration)  acts to reduce the OP, possibly down to a lower value than 
the one expected from bandstructure calculations neglecting strong correlation effects. 
These issues were also considered at the model level in Ref.~\onlinecite{Parragh2013}. 

On the experimental side, x-ray absorption (XAS) and x-ray linear dichroism (XLD) 
spectroscopy, combined with resonant reflectivity, were recently performed\cite{Wu2013} for a series 
of LNO heterostructures under a wide range of strains from $-2.3$\% to $+3.2$\%. 
These data clearly revealed that an orbital polarization $P=(n_{\xxyy} - n_{\zz})/(n_{\xxyy}+n_{\zz})$ as high as 
20-25\% is achieved under tensile strain, with an essentially linear dependence of the polarization on strain. 
Note however that the results of another experimental investigation on thin films were interpreted 
as the occurrence of an octahedral breathing-mode distortion under tensile strain 
with negligible orbital polarization.\cite{Chakhalian2011,Blanca-Romero2011}

In this work, we investigate how strain affects the orbital degrees of freedom 
in \lanio epitaxial films. We take into account both the effect of strain-induced 
structural distortions and electronic correlations using many-body 
electronic structure methods. 
The calculated values of the orbital polarization as a function of strain are in good agreement 
with the experimental results of Ref.~\onlinecite{Wu2013}. 
We also find that, for realistic values of interaction parameters, 
correlation effects associated with the Hund's rule coupling reduce 
the strain-induced  polarization as compared to the value obtained from 
band-structure calculations, 
in agreement with the conclusions of previous works.\cite{Han2011}

This paper is organized as follows. To begin with, we consider the effects of
strain on the crystal structure of \lanio in Sec.~\ref{sec:struct}. 
Then we discuss the strain-induced orbital polarization, first from a 
bandstructure standpoint (Sec.~\ref{sec:op_gga}) and then including electronic 
correlation effects in Sec.~\ref{sec:dmft}.  Finally, we compare 
our theoretical calculations to experimental results and
discuss in some details interpretations of the latter in Sec.~\ref{sec:comp}. 
Readers mostly interested in the final results
may jump to this last section and, in particular,
to Figs.~\ref{fig:op_exp},~\ref{fig:rhole}.
Also, some additional details can be found in Appendixes A and B.

%
%
\section{Effects of strain on the structure of \lno}
\label{sec:struct}

Most of the rare-earth (RE) nickelates possess a perovskite
ABO$_{3}$ structure with various
distortions depending on the temperature and composition.\cite{Medarde1997, Catalan2008}
In particular, the difference in ionic radii and the mismatch of
the B-O and A-O equilibrium bond lengths $d_{XO}$
result in octahedral tilts whose magnitude can be related to
the tolerance ratio, $t = d_{AO}/d_{BO}\sqrt{2}$, quantifying  
the deviation of the bond-length ratio from the one of the ideal perovskite structure.
The octahedral tilts lead to the decrease of Ni-O-Ni bond
angles and the reduction of the Ni$d$-O$p$ hybridization, which, in turn,
has a significant impact on the electronic structure.
This effect is manifested in a direct dependence of
the temperature of the metal-insulator transition (MIT) on $t$,
with smaller $t$ leading to higher critical temperatures.\cite{Torrance1992}
Bulk \lanio, having the largest value of $t$, is the only compound in the
family of RE nickelates that remains metallic and does not undergo a
transition down to lowest temperatures.

Bulk \lanio has a perovskite structure with a rhombohedral distortion of
the unit cell and the corresponding space group is $R\bar{3}c$.\cite{Garcia-Munoz1992} 
The rhombohedral distortion is induced by rotations of Ni-O octahedra
with a rotation pattern of type $a^{-}a^{-}a^{-}$
(in Glazer notation \cite{Glazer1972,Glazer1975}), which means that rotations around all
three axes are anti-phase (the direction of the rotation around an axis
is alternating along the given axis).
This is in contrast to other nickelates having rotation pattern
$a^{-}a^{-}c^{+}$ (GdFeO$_{3}$ type), i.e. with rotations around the $a$ and $b$
axes being anti-phase and the rotation around the $c$ axis in-phase.

The aim of this section is to determine how strain affects the structure of 
\lno, as compared to the unstrained bulk. 

\subsection{Setup and Method}
We consider a layer of \lanio on a cubic or tetragonal substrate with a
square in-plane face of the pseudo-cubic cell 
(as in STO or LSAT substrates).\footnote{For such substrates all four possible
orientations of the LNO trigonal rotation axis with respect to the 
substrate plane are equivalent.\cite{May2010}}
A mismatch of the equilibrium lattice parameter
between LNO and the substrate causes the LNO layer to be subject to
biaxial strain in the $ab$-plane. In nickelate films the strain can be
sustained within rather thick films up to 20--50 atomic layers 
(depending on strain),\cite{Scherwitzl2009} 
and we can thus neglect film-substrate interfacial as well as surface effects. 
The effect of the strain reduces then to a geometrical constraint on the
bottom in-plane face of the pseudo-cubic cell of 
LNO.~\footnote{It is worth noting that since bulk LNO has a rhombohedrally distorted
unit cell, a tetragonal substrate exerts a small \textit{axial} strain
(along $a$ or $b$ axis) on the film even when the biaxial strain is zero
(as defined from the ratio of the lattice constants).}
Hence, a sufficiently thick film can be modeled by a bulk-like geometry 
in which the in-plane lattice parameters (as well as the angle between the in-plane vectors)
are fixed to those of the substrate and all other degrees of freedom are allowed to relax.

To identify the crystal structure of a strained film we perform structure optimization
within generalized gradient approximation (GGA) \cite{pbe96} using
the projected-augmented waves (PAW) method \cite{paw94} as implemented in
the Vienna \textit{ab-initio} simulation package (VASP). \cite{paw_vasp,vasp1,vasp2}
The integration over the Brillouin zone is done using a
$k$-mesh with $11\times 11\times 11$ points and the plane-wave
cutoff is $E_{\textrm{cut}} = 600$ eV. Structure relaxation is considered
to be converged when the forces are smaller than $10^{-3}$~eV/\AA.

To simulate the LNO layer we set up a base-centered 
monoclinic unit cell (space group $C2/c$) oriented in such a way
that Ni positions match those of B-ions of a [001]-oriented ABO$_{3}$ cubic substrate.
This crystal space group is determined by the pattern of octahedral rotations
(see Ref.~\onlinecite{Glazer1972})
and epitaxial constraints and was experimentally found in LNO/LAO and LNO/STO films.\cite{May2010} 
The in-plane parameters of
the pseudocubic cell, $a_{p} = b_{p}$, are fixed to that of the substrate, which
determines the strain $\epsilon_{xx} \equiv a_{p} / a_{p,\textrm{eq}} - 1$,
where $a_{p,\textrm{eq}}$ is the pseudocubic lattice parameter of
bulk LNO ($a_{p,\textrm{eq}} = 3.863$ as obtained within GGA).
All other degrees of freedom, such as the out-of-plane
lattice vector, oxygen and cation positions, are allowed to relax.

It is also instructive to compare the case of a relaxed monoclinic unit-cell 
with the situation in which the system is constrained to remain tetragonal 
with no octahedral rotations (relaxing only $c/a$). Indeed, this
reveals the role of the rotations and unit cell monoclinic distortion
in the structural response to strain. 
In the following, we shall refer to the monoclinic and tetragonally-constrained cases
as `\dist' and `\tetra', respectively.

%
\begin{figure}
\includegraphics[width=1.0\linewidth]{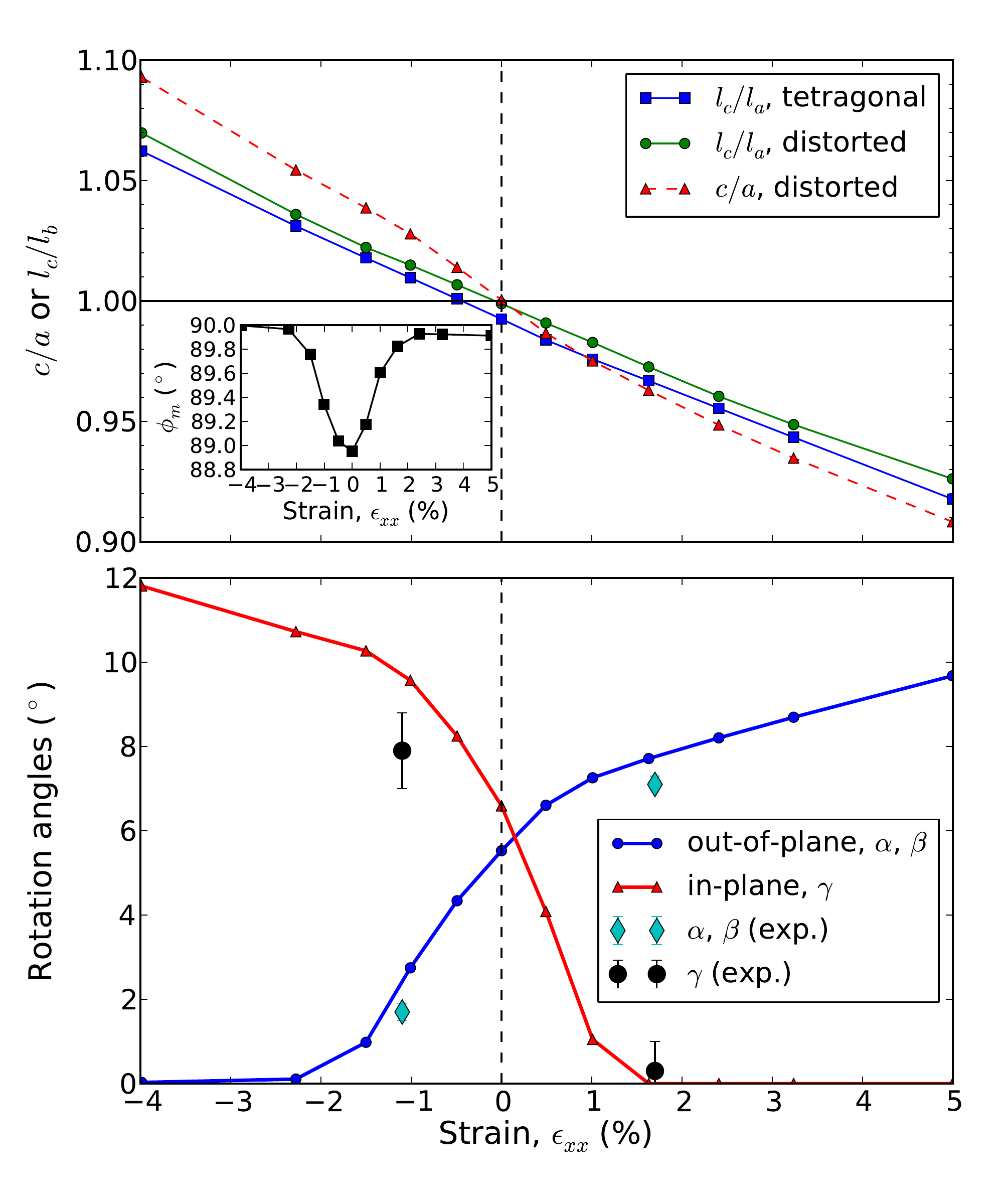}
\caption{(Color online) Top: Bond-length ratio, $l_{c}/l_{a}$, for the distorted (solid green) and
tetragonal (solid blue) structures under strain.
The broken line displays the $c/a$ ratio for the distorted structure 
(for the tetragonal structure it is identical to the bond-length ratio).
In both cases, the strain is defined with respect to $a_{p,\textrm{eq}}$
of bulk LNO; the shift of the zero-strain point in the tetragonal case
reflects thus the difference in the lattice constants of the two types of
structures.
Bottom: Dependence of the octahedral in-plane rotations ($\gamma$)
and out-of-plane tilts ($\alpha=\beta$) on strain for the
fully relaxed \dist structure. Also, structural refinement data from Ref.~\onlinecite{May2010}
are shown with diamonds for $\alpha=\beta$ and with circles for $\gamma$.
Inset: Inclination angle $\phi_{m}$ of the pseudocubic
axis $c_{p}$ with respect to the $ab$ plane.
}
\label{fig:coa}
\end{figure}
%

\subsection{Results}
The first important effect of strain is the contraction/expansion 
(for tensile/compressive strain, respectively) of
the unit cell in $z$-direction. This is measured by the $c/a \equiv c_{p}/a_{p}$ ratio, 
or equivalently, by the $z$-strain $\epsilon_{zz} \equiv c_{p}/c_{p,\textrm{eq}} - 1$,
where $c_{p}$ is the length of the out-of-plane axis of the pseudocubic unit cell. 

We display in the top panel of Fig.~\ref{fig:coa} the results for $c/a$ as
a function of strain for both the 
\tetra and \dist cases. 
If one compares the evolution of the $c/a$ ratio for the \tetra case, in which
the dependence is almost perfectly linear, to that of the \dist case, 
one can see that the octahedra tilts result in a noticeably
non-linear elastic response.
At the same time, the ratio of the octahedron bond lengths, $l_{c}/l_{a}$, also displayed 
in Fig.~\ref{fig:coa}, is very similar in both cases, implying that the internal geometry
of the octahedra is similar for the \tetra and \dist structures.  
The non-linear behavior of the $c/a$ ratio in the \dist case reflects
thus the evolution of the rotation angles under strain.

Another important structural effect, with direct consequences for the electronic structure, 
is the different pattern of octahedral rotations found for compressive and tensile strains
(the bottom panel of Fig.~\ref{fig:coa}), 
as previously discussed in Ref.~\onlinecite{May2010}.  
Octahedral rotations can be characterized by angles 
$\alpha$, $\beta$, $\gamma$ of rotations around the $x,y$-axis (out-of-plane rotations) 
and $z$ axis (in-plane), respectively.\cite{Glazer1972} 
In unstrained bulk \lno, all three angles are equal $\alpha=\beta=\gamma$, and the 
system has a rhombohedral symmetry. 
Under compressive strain, out-of-plane rotations (tilts) are suppressed and
at the most negative strain only the in-plane rotation is left,
with the structure approaching a higher tetragonal symmetry.
The system under tensile strain, on the other hand, prefers 
octahedra to tilt, with the in-plane rotations being suppressed already
at moderate strains. For strains corresponding to LaAlO$_{3}$ (LAO) and
STO substrates, the angles are in good agreement with the available
experimental data,\cite{May2010} as also shown in Fig.~\ref{fig:coa}.
Note, however, that tilt angles in
superlattices might differ from those in films. \cite{Hwang2013}

This behavior can be considered as a second-order structural isosymmetric transition
that the LNO layer undergoes on crossing over from compressive to tensile strain, 
whereby the rotation pattern
of octahedra changes from $a^{0}a^{0}c^{-}$ (in-plane rotation $\gamma\neq 0$, 
$\alpha=\beta\simeq 0$)  to $a^{-}a^{-}c^{0}$ (out-of-plane tilting 
$\alpha=\beta\neq 0, \gamma\simeq 0$). 

The rigidity of octahedra dictates that anti-phase rotations around three axes in bulk \lanio
must induce distortions of the unit cell, with the pseudocubic vectors 
$a_{p}$, $b_{p}$, $c_{p}$ being inclined with respect to each other.\cite{Glazer1972}
In an epitaxially constrained film, the angles between $a_{p}$ and $b_{p}$ are
fixed by the substrate and only the $c_{p}$ axis can relax in such a way as to
avoid a strong deformation of the octahedra. This is perfectly illustrated by
the dependence of the inclination angle $\phi_{m}$ of $c_{p}$ with respect
to the $ab$ plane (inset in Fig.~\ref{fig:coa}). The largest deviation from
90$^{\circ}$ is taking place around zero strain, which is important for
stabilizing the configuration with all angles being equal ($a^{-}a^{-}a^{-}$, as in the bulk).
Indeed, when constraining $c_{p}$ to be orthogonal to the plane, GGA calculations (not shown)  
yield a very sharp first-order-like transition, with 
the $a^{-}a^{-}a^{-}$ configuration being unstable at zero strain. This is 
not consistent with the bulk crystal structure, and emphasizes the importance of 
letting the structure fully relax in the calculations. 

Finally, we would like to mention that earlier works relying on GGA+U calculations
predicted that \lanio films experience bond disproportionation under 
tensile strain.\cite{May2010,Chakhalian2011} However, these calculations 
assume some form of magnetic and/or orbital ordering. Recent more precise results obtained 
within the DFT+DMFT approach indicate
that GGA+U overestimates this effect,\cite{Park2014} and we therefore 
do not expect that the bond-disproportionated phase is relevant for the range of 
strains considered in the present paper.
This is also supported by recent experimental results.\cite{Wu2013}

\section{Orbital polarization: Effect of structural distortions}
\label{sec:op_gga}
In this section, we investigate the effects of structural distortions on the orbital 
polarization of $e_g$ states in strained LNO. All calculations are performed within GGA:  
the effects of electronic correlations in the Ni $d$-shell, and their interplay with structural aspects 
will be considered in the next section. 

\begin{figure}
\includegraphics[width=1.0\linewidth]{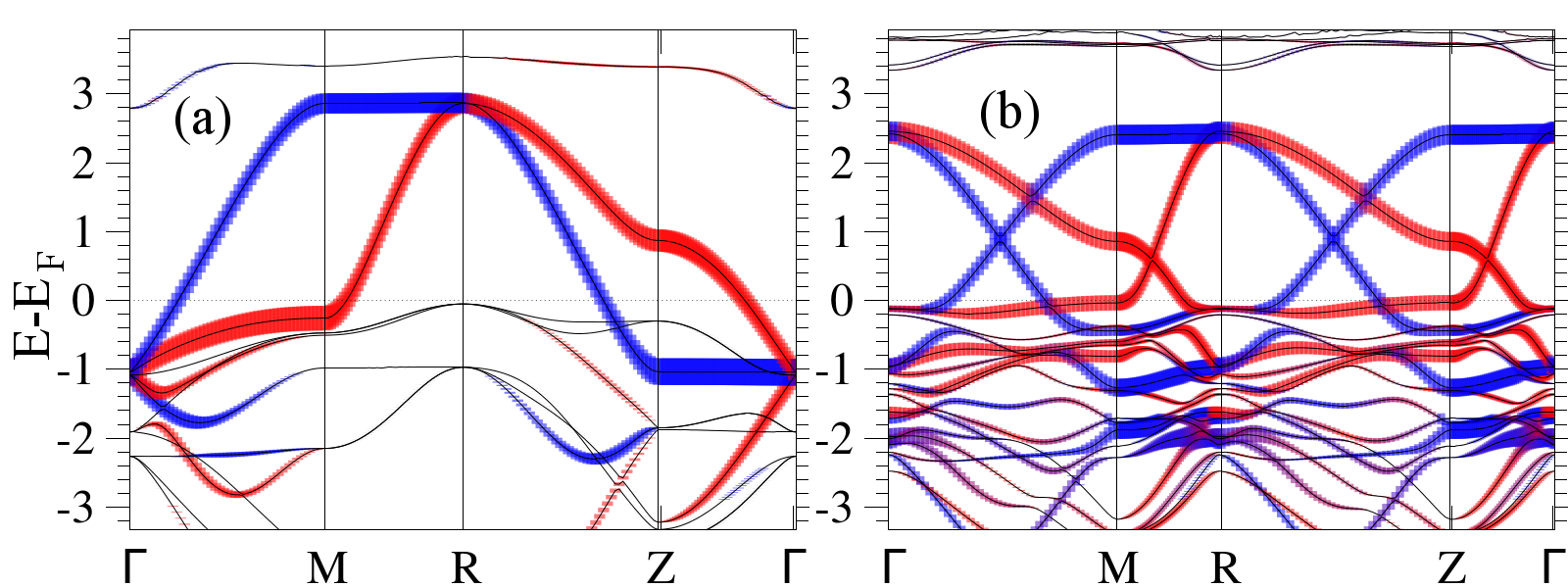}
\caption{(Color online) Band structure at zero strain (GGA):  
(a) \tetra structure, (b) \dist structure. 
A `fat-band' representation is used to indicate orbital character: 
red (dark) corresponds to $d_{\zz}$, blue (light) to $d_{\xxyy}$ orbitals. 
A Brillouin zone of the cubic cell is used in all cases.
A clear flattening of bands ($\Gamma$-$M$) around the Fermi level is observed for the 
\dist structure, as a result of hybridization with the $t_{2g}$ states.}
\label{fig:bs_unstrained}
\end{figure}

We start with a brief review of the basic electronic structure of LNO. 
The formal valency of Ni-ions
is $3^{+}$ ($d^7$), with the ionic ground-state configuration being $t_{2g}^{6}\eg^{1}$.
This is only a formal assignment however, since strong covalency and hybridization with oxygen states
usually lead to a nominal valency different from the formal one.
The GGA band structure of unstrained LNO in both the \tetra and \dist structures is displayed in 
Fig.~\ref{fig:bs_unstrained}.
The orbital characters in the local frame of tilted octahedra are obtained
by real-space rotation of the basis by tilt angles.
The $t_{2g}$-like bands lie
below the Fermi level but are quite close in energy. 
They are completely filled, and their dispersion is relatively weak at the top of the bands~\cite{Eguchi2009}. 
The behavior of valence electrons is thus almost entirely
determined by the $\eg$ states and the oxygen $p$ states hybridized with the $\eg$ states, 
i.e., the `$e_g$-like' bands. 
In Fig.~\ref{fig:bs_unstrained}, we also see that in the \dist structure, the bands acquire a more mixed 
orbital character 
(away from $\Gamma$ point)
in terms of the orbitals defined
in the local reference frame of the (tilted) octahedra.
As a result, the relevant bands near the
Fermi level have a sizable $t_{2g}$ contribution as well.
\footnote{Note that the mixing of bands happens only at non-zero
$k$ vectors. The orbital characters at $\Gamma$ point are entirely determined
by the local crystal-field symmetry.}

\begin{figure}
\includegraphics[width=1.0\linewidth]{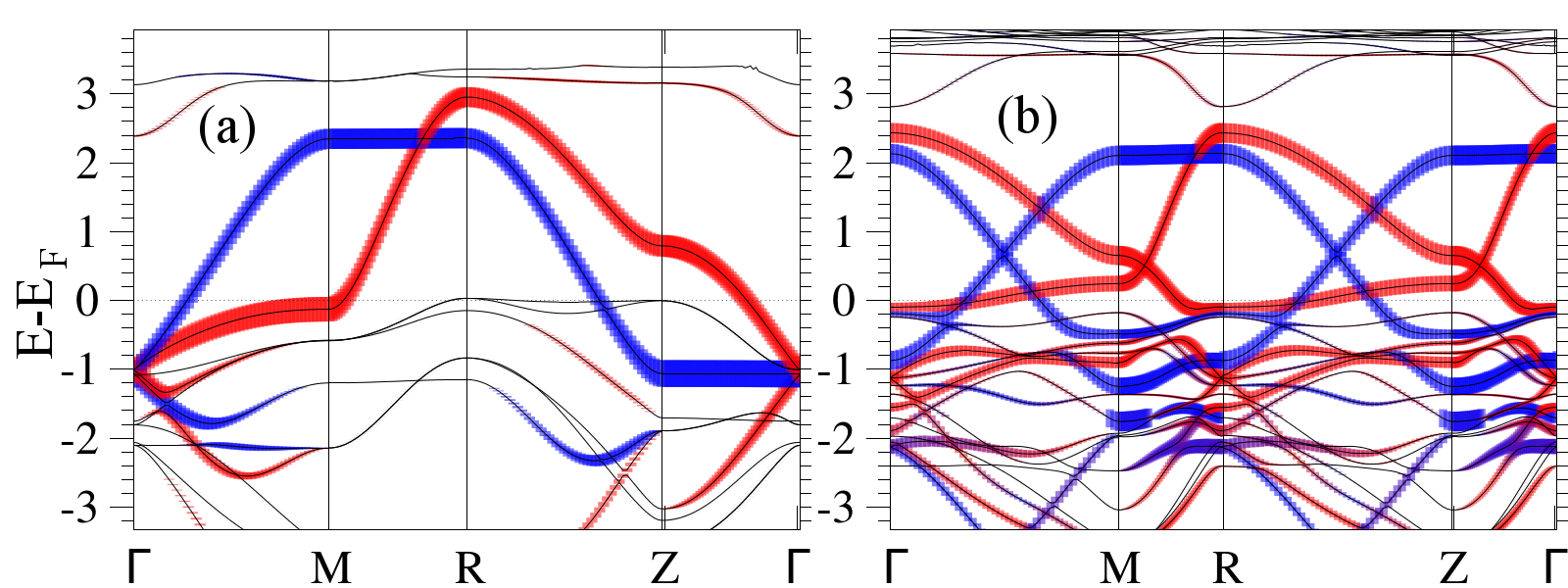}
\caption{(Color online) Band structure (GGA) for a $+3.2\%$ tensile strain:  
(a) \tetra structure, (b) \dist structure. 
A `fat-band' representation is used to indicate orbital character: 
red (dark) corresponds to $d_{\zz}$, blue (light) to $d_{\xxyy}$ orbitals. 
The center of gravity of the $d_{\zz}$ is pushed upwards, and the bandwidth of the 
$d_{\xxyy}$ band is reduced. Note that the two bands remain degenerate at the $\Gamma$ point.}
\label{fig:bs_strained}
\end{figure}

In the presence of epitaxial strain, the changes in the structure induce a lifting of the degeneracy of the $\eg$ states. 
This is clear from Fig.~\ref{fig:bs_strained} which displays the GGA band structure under a $+3.2\%$ tensile strain. 
As expected, due to the compression of the octahedra, the center of gravity of the $d_{\zz}$ band is pushed upwards relative 
to that of the $d_{\xxyy}$ band.
The bandwidth of the $d_{\xxyy}$ band is also reduced (due to the larger in-plane lattice constant). 
The two bands, however, remain almost completely degenerate at the $\Gamma$ point
due to vanishing $d$-$p$ hybridization at this point.
\footnote{Lifting of the degeneracy at the $\Gamma$ point can be
achieved by controlling the direct hoppings (not mediated by oxygens) between Ni $d$ states
in plane and perpendicular to the plane.\cite{Hansmann2009}
In particular, hoppings are anisotropic in other members of Ruddlesden-Popper series (e.g., A$_{2}$BO$_{4}$
as in cuprates) or in heterostructures.}
Importantly, the described properties are still valid even when correlations
are taken into account (see Appendix B).
%
The degeneracy lifting of the $e_g$ states can be quantified by defining 
the intra-$\eg$ crystal-field (CF) splitting as:
\begin{equation}
\Delta_{e_g}\equiv \eps_{\zz} - \eps_{\xxyy},
\end{equation} 
where the level positions $\eps_{\zz}$, $\eps_{\xxyy}$ are given by the
diagonal terms of the local Hamiltonian (obtained by projecting the
Kohn-Sham Hamiltonian onto the $\eg$ states), which is equivalent to finding the
center of mass of the projected DOS for each orbital.

This results in a corresponding change in the relative orbital occupancies $n_{\xxyy}$, $n_{\zz}$, 
leading to OP quantified as
\begin{equation}
\pol \equiv \frac{n_{\xxyy} - n_{\zz}}{n_{\xxyy} + n_{\zz}} = 
\frac{1}{n_{\eg}}(n_{\xxyy} - n_{\zz}),
\label{eq:op}
\end{equation}
which is the quantity of central interest in this paper.
An important question is the proper definition of the occupancies 
$n_{\xxyy},n_{\zz},n_{e_g}=n_{\xxyy}+n_{\zz}$ entering 
this expression. We now address this question in some detail. 

%
%
%
%

\subsection{Definition of occupancies 
\label{sec:basis}}
Strong $d$-$p$ hybridization in transition-metal oxides makes
orbital occupancies (such as $n_{\zz}$, $n_{\xxyy}$)
quite sensitive to the choice of the local basis\cite{Mizokawa2000,Parragh2013}
with respect to which these occupancies are calculated.
This is particularly pronounced in systems with a small charge-transfer
energy, such as nickelates, in which the orbital character of states
close to the Fermi level is determined by a mixture of $d^{7}$ and
$\dlbar$ states,\cite{Abbate2002,Liu2011,Mizokawa1994,Moon2012}
each of these states obviously having a different $d$-electron count.

There are essentially two strategies to define the local-basis
Wannier functions differing by the choice of the
energy window in which the projection is done 
, or, equivalently, by the choice of the subset of projected bands.
The first way is to choose a broad energy window involving
all Ni $d$-like and oxygen $p$-like bands (for \lanio this corresponds
to bands within a range $[-8.0, 3.0]$ eV).
The resulting localized Wannier (\bw{}) functions, $\ket{\chi}_{\bw}$,
are well localized, have predominantly $d$-character and yield
the total occupation of $\eg$ states close to 
$n_{\eg} \approx 2$ [\onlinecite{Deng2012}]. 
On the other hand, choosing a narrow energy window ($[-1.6, 3.0]$ eV)
embracing mainly the $\eg$-like bands around the Fermi level
leads to extended Wannier (\sw{}) functions $\ket{\chi}_{\sw}$
with substantial weight on the oxygen sites. 
These extended Wannier functions comprise both the localized Ni$d$-states
and a linear combination of the O$p$ states
having the same symmetry as the $\eg$ states ($p_{\sigma}$-states).
The total occupation of $\eg$ states is $n_{\eg} \approx 1$ in this case.

In this paper we will be using the \sw{}-type of basis 
for the evaluation of the local quantities, including the OP and CF splitting.
There are two main reasons justifying this choice, 
as discussed in more details below:  
(i) this choice ensures that the calculated OP provides information about 
the degree of orbital polarization of low-energy quasi-particle bands and 
(ii) importantly, the OP defined in this manner is consistent with the 
one measured in XAS and resonant spectroscopy experiments.

Regarding (i), it is natural to define the OP in such a way that 
a system reaching $P = 100\%$ is associated with a transition from an initially
two-sheet ($\eg$-like) Fermi surface to a single-sheet one
($d_{\xxyy}$-like) when a sufficiently large CF splitting is imposed.
In this case, the $d_{\zz}$-band becomes completely unoccupied
and the system becomes effectively single-band.
The orbital character of these low-energy bands is
primarily determined by the symmetry of the local states,
the latter being represented by both the Ni$d$- and O$p_{\sigma}$ states.
Such states are consistently described by \sw{}-type
Wannier functions, $\ket{\chi_{\eg}}_{\sw}$.

From the experimental standpoint (ii), spectroscopic probes 
such as linear dichroism in XAS and resonant spectroscopies\cite{Benckiser2011} are by design
sensitive to the symmetry of the local states.
When one wants to extract
$P$ from the results of a dichroism measurement,
one has to make assumptions about the overlap
between the local exciton and both the $d$ states of nickel
and the symmetrized oxygen $p_{\sigma}$ states.
Since the latter penetrate quite deep inside Ni-O octahedra,
it is natural to expect a substantial contribution
to the OP from the oxygen hole states, which is well captured
by the extended \sw{} basis. In other words, we assume that XAS does not
promote a core electron to a very localized atomic-like Ni state
but rather to extended states of mixed Ni$d$-O$p$ character.

\subsection{Results and Analysis}

\begin{figure}
\includegraphics[width=1.0\linewidth]{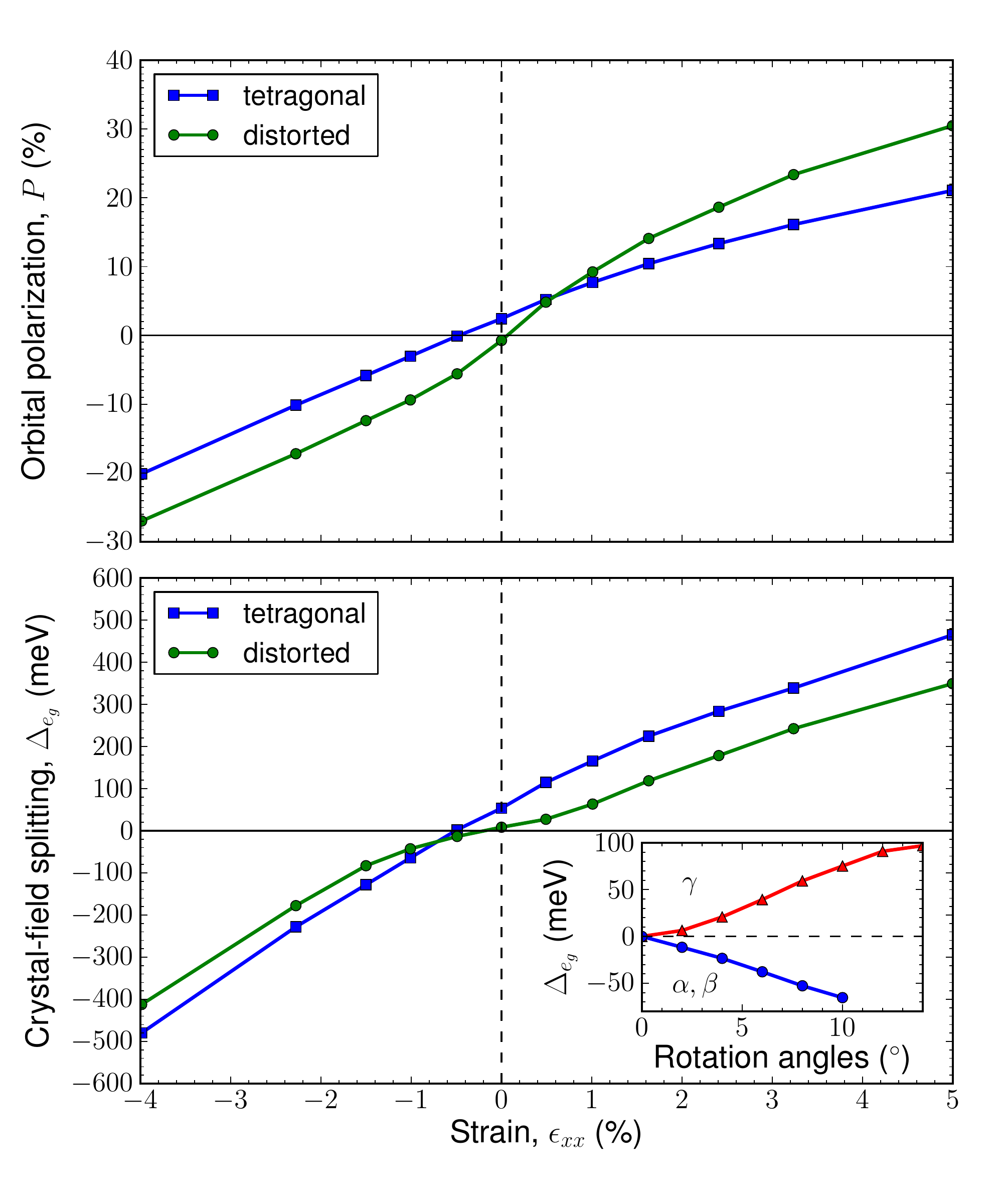}
\caption{(Color online) Top: The orbital polarization, $P$, obtained within GGA 
as a function of strain for the
tetragonal (squares, blue) and distorted structures (circles, green).
(Color online) Bottom: The corresponding $\eg$ crystal-field splitting, $\Delta_{\eg}$, 
as a function of strain for the
tetragonal (squares, blue) and distorted structures (circles, green).
Inset: $\Delta_{\eg}$ as a function of in-plane and out-of-plane rotation
angles ($\gamma$ and $\alpha$,$\beta$, respectively), with the $l_{c}/l_{a}$ ratio
fixed to 1.
}
\label{fig:op_gga}
\end{figure}

The top panel of Fig.~\ref{fig:op_gga} displays the orbital polarization
as a function of strain obtained 
within GGA. One can see a substantial difference in the behavior of the OP
in the \tetra and \dist structures. In the case of
the tetragonal structure, the dependence of $P$ on strain is smooth 
and lacking any features, as expected from a simple
picture where the polarization is induced by a uniform relative shift
of the $d_{\zz}$-like and $d_{\xxyy}$-like bands. 
In contrast, in the \dist structure the OP is significantly enhanced
over the entire range, except for a small region around zero strain where
the plot reveals a transitional behavior when going over from compressive
to tensile strain. At the same time, the slope of the OP is
almost the same at large strains for both types of structure.
This implies that the enhancement is due to the electronic
structure of the system at small strain.

To understand the cause of the polarization enhancement in
the \dist structure as compared to the tetragonal one, we
plot the intra-$\eg$ CF splitting, $\Delta_{\eg}$, as a function of strain in 
the bottom panel of Fig.~\ref{fig:op_gga}. 
In contrast to the OP, we find that the CF splitting is {\it smaller} in the \dist case. 
This is expected qualitatively: allowing the structure to relax (mostly by tilting the 
octahedra) will indeed alleviate the effect of the strain imposed on the structure and 
reduce the CF splitting. 

In order to disentangle the effects of octahedral tilts on $\Delta_{\eg}$ 
from that of the change in bond length ratio $l_{c}/l_{a}$,
we have calculated the crystal field as a function of rotation angles $\gamma$ and
$\alpha$,$\beta$, with the ratio $l_{c}/l_{a}$ being kept fixed
(inset of Fig.~\ref{fig:op_gga}).
Having in mind the dependence of rotation angles on strain [Fig.~\ref{fig:coa}(b)],
it is clear that octahedral rotations tend to
reduce the absolute value of the crystal-field splitting induced by strain.

\begin{figure}
\includegraphics[width=1.0\linewidth]{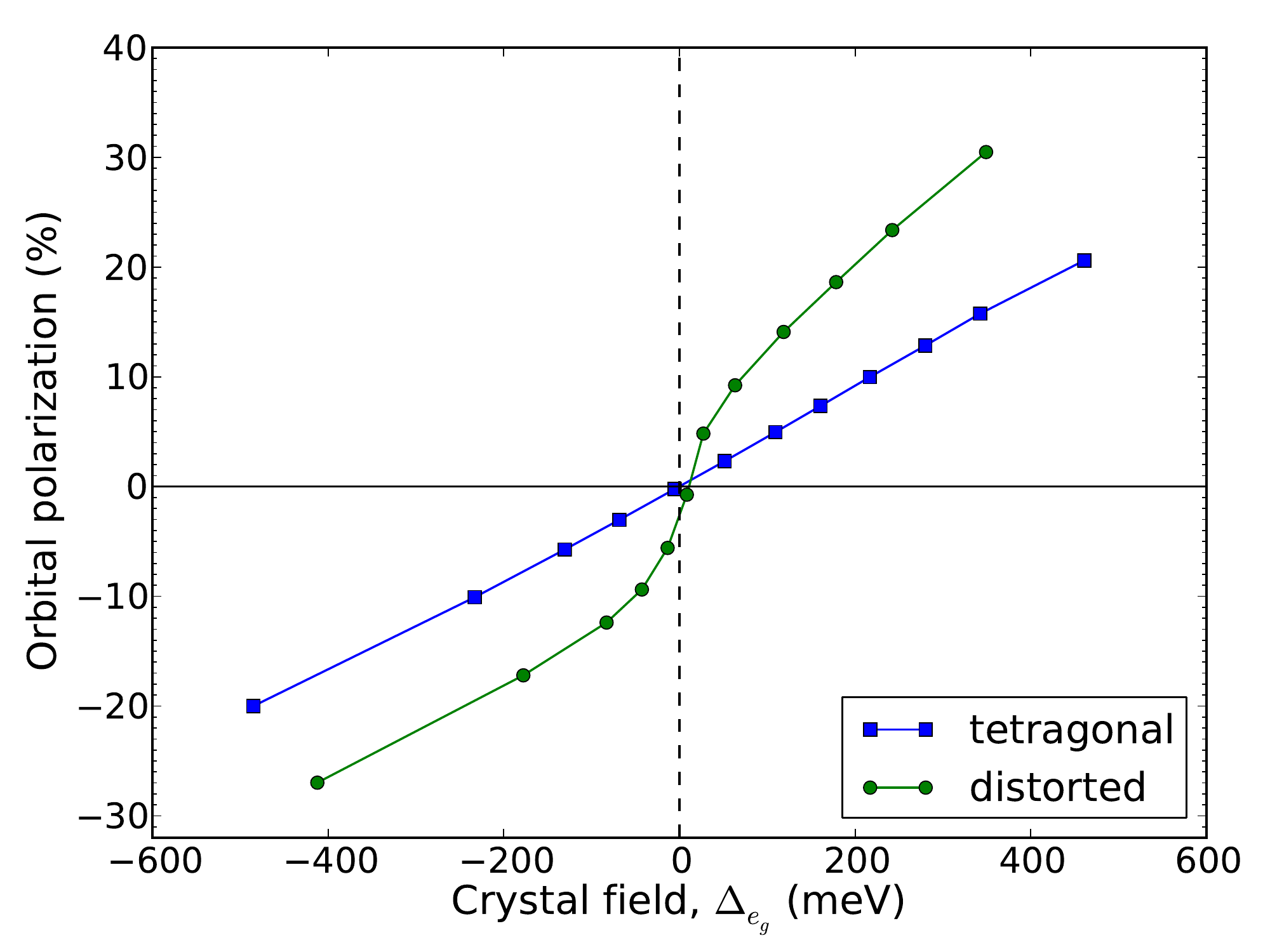}
\caption{(Color online) Orbital polarization $P$ obtained within GGA as a function of the crystal-field
splitting, $\Delta_{\eg}$, for the 
\tetra (squares, blue) and \dist structures (circles, green).
}
\label{fig:op_cf}
\end{figure}

Finally, we display in Fig.~\ref{fig:op_cf} the OP as a function of 
CF for each structure. We see that while the \tetra structure displays 
an almost perfect linear dependence of $P$ on $\Delta_{\eg}$ this dependence 
has a sharp critical-like behavior for the \dist structure.
This reflects the non-linear feedback of 
octahedral rotation and tilts when the structure is relaxed. 
%
%
%
%

Defining the orbital polarizability $\psusc$ at small strain from the slope of the 
dependence of the OP vs. CF: 
\begin{equation}
\pol = \psusc\,\Delta_{\eg},
\end{equation}
we see that $\psusc$ is almost four times larger for the \dist than for the \tetra structure.
In a simple rigid-band model in which the crystal field simply shifts the partial 
density of states of the $d_{\zz}$ and $d_{\xxyy}$ orbitals (denoted by 
$D_z$ and $D_x$, respectively), the orbital susceptibility is easily obtained as: 
\begin{equation}
\psusc = 2 \frac{D_{x} D_{z}}{D_{x} + D_{z}} = D_{\eg},
\label{eq:op_rigid}
\end{equation}
in which all density of states are taken at the common Fermi level. 
The last relation holds when the orbitals are degenerate at
zero strain so that $D_{x} = D_{z} \equiv D_{\eg}$. 

\begin{figure}
\includegraphics[width=0.9\linewidth]{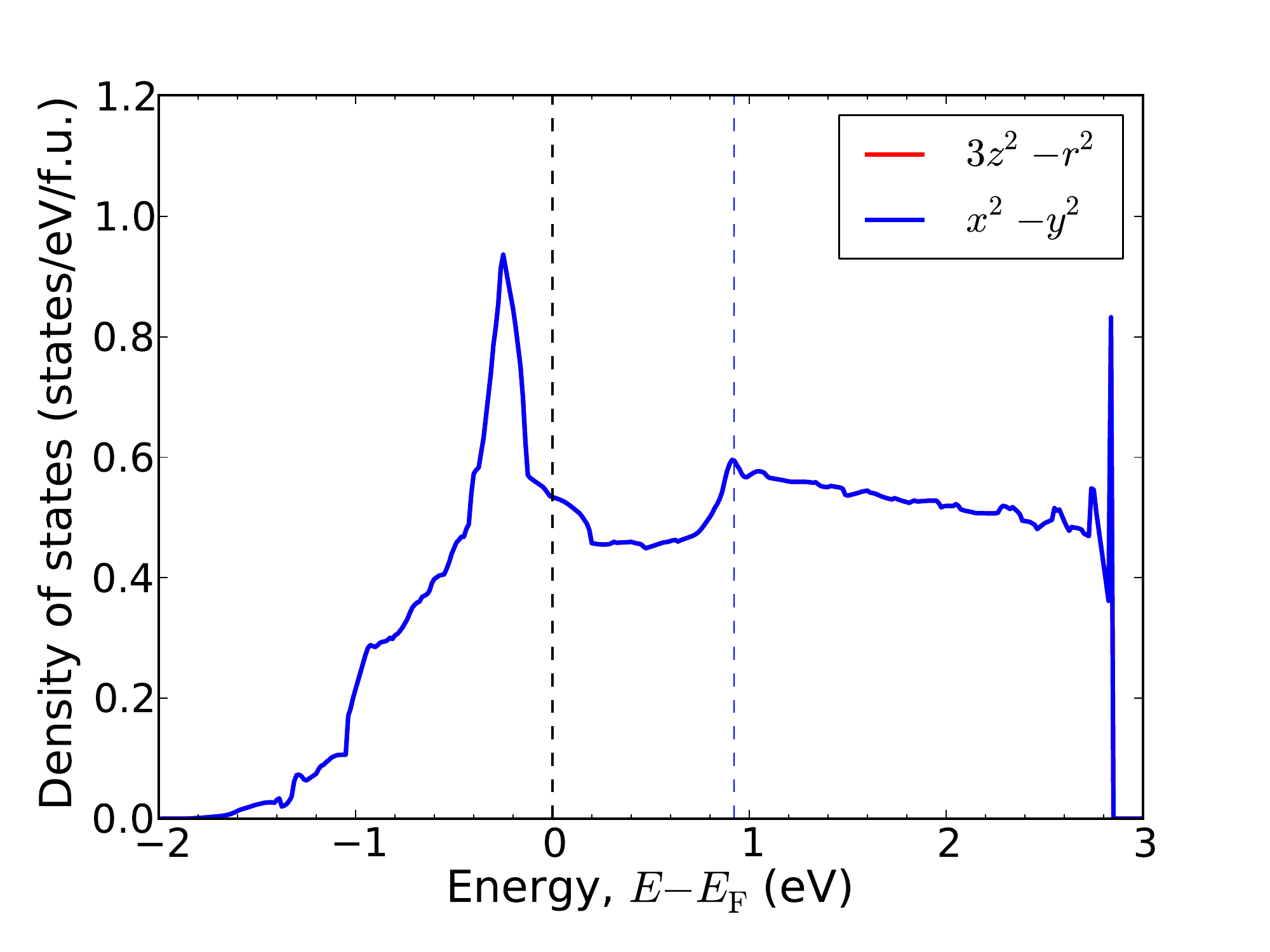}
\caption{(Color online) Partial $\eg$ DOS of the tetragonal structure with $c/a = 1$.}
\label{fig:dos_tetra}
\end{figure}

\begin{figure}
\includegraphics[width=0.9\linewidth]{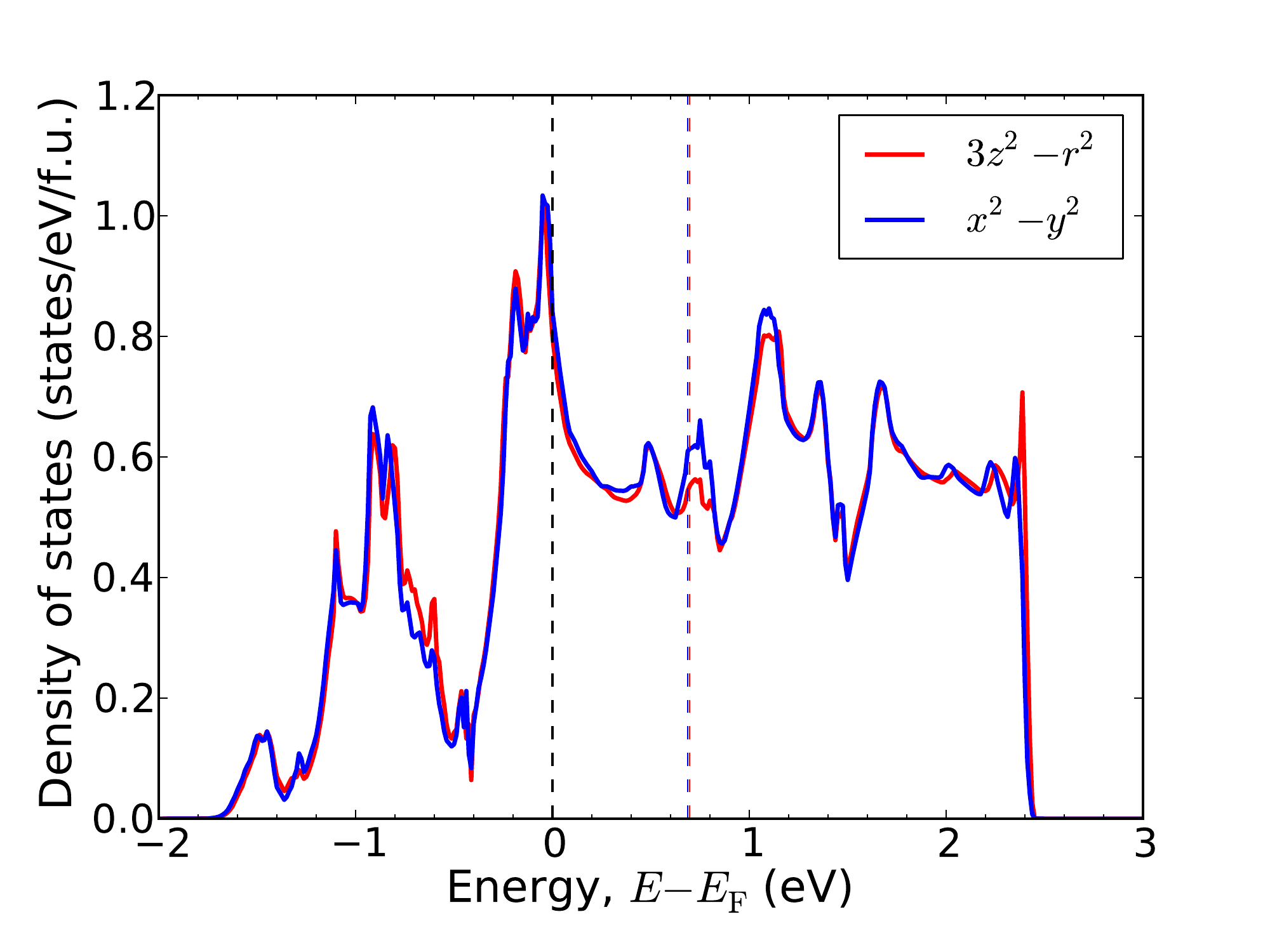}
\caption{(Color online) Partial $\eg$ DOS of the distorted structure at zero strain. Keep in mind
that this structure is slightly different from the true bulk structure
with a higher rhombohedral symmetry.}
\label{fig:dos_relax}
\end{figure}

The DOS for the \tetra and \dist structures are displayed in Figs.~\ref{fig:dos_tetra} and 
\ref{fig:dos_relax}, respectively. Remarkably, the Fermi level of the \dist 
structure falls close to a sharp peak in the DOS, which is not the case for 
the \tetra structure for which the Fermi level falls in a featureless flat region. 
This explains qualitatively the larger orbital susceptibility of the 
\dist structure. On a quantitative level, the value of $\psusc$ is in 
rather good agreement with the calculated DOS $D_{\eg} = 0.53$ eV$^{-1}$ 
and expression (\ref{eq:op_rigid}) for the \tetra case. 
In contrast the value of the $D_{\eg} = 0.82$ eV$^{-1}$, $1.6$ times larger 
than for the \tetra structure, is smaller than the calculated large 
value of the orbital polarizability enhancement (Fig.~\ref{fig:op_cf}). 
This is because a rigid-band model does not properly take into account the 
non-linearities associated with relaxation and the intrinsic changes in the DOS 
under strain. 

The origin of the peak in the DOS for the \dist structure can be understood by 
looking at the bandstructure in Fig.~\ref{fig:bs_unstrained}. 
Octahedral tilts lead to hybridization between the $\eg$ states and relatively flat $t_{2g}$ bands. 
The corresponding flattening of $\eg$ bands near the Fermi level is clearly observed
on the projected band structure (Fig.~\ref{fig:bs_unstrained}).
A similar enhancement of the DOS at the Fermi level was reported in earlier works.\cite{Hamada1993}

We have demonstrated that the observed larger orbital polarizability of 
the \dist structure as compared to the \tetra structure is due to differences 
in the electronic structure, and especially due to the hybridization with $t_{2g}$ bands 
when octahedra relax and tilt. In the next section we consider how the
OP is affected by electron correlations.

Finally, it is worth noting that epitaxial strain alone is inefficient
in lifting the degeneracy of $\eg$ orbitals at the $\Gamma$ point [compare, e.g.,
Figs.~\ref{fig:bs_unstrained}(a) and \ref{fig:bs_strained}(a)], which limits
its ability to produce very large OP.\cite{Hansmann2009,Yang2010}
The efficiency can be increased by sandwiching single nickelate layers between
insulating layers in a superlattice, confining thus carriers inside the plane
and lifting the $d_{\zz}$ band with respect to the $d_{\xxyy}$ band.\cite{Han2010,Chen2013b}

%
%
%
%

\section{Orbital polarization: Effect of correlations\label{sec:dmft}}
%
%
\begin{figure}
\includegraphics[width=1.0\linewidth]{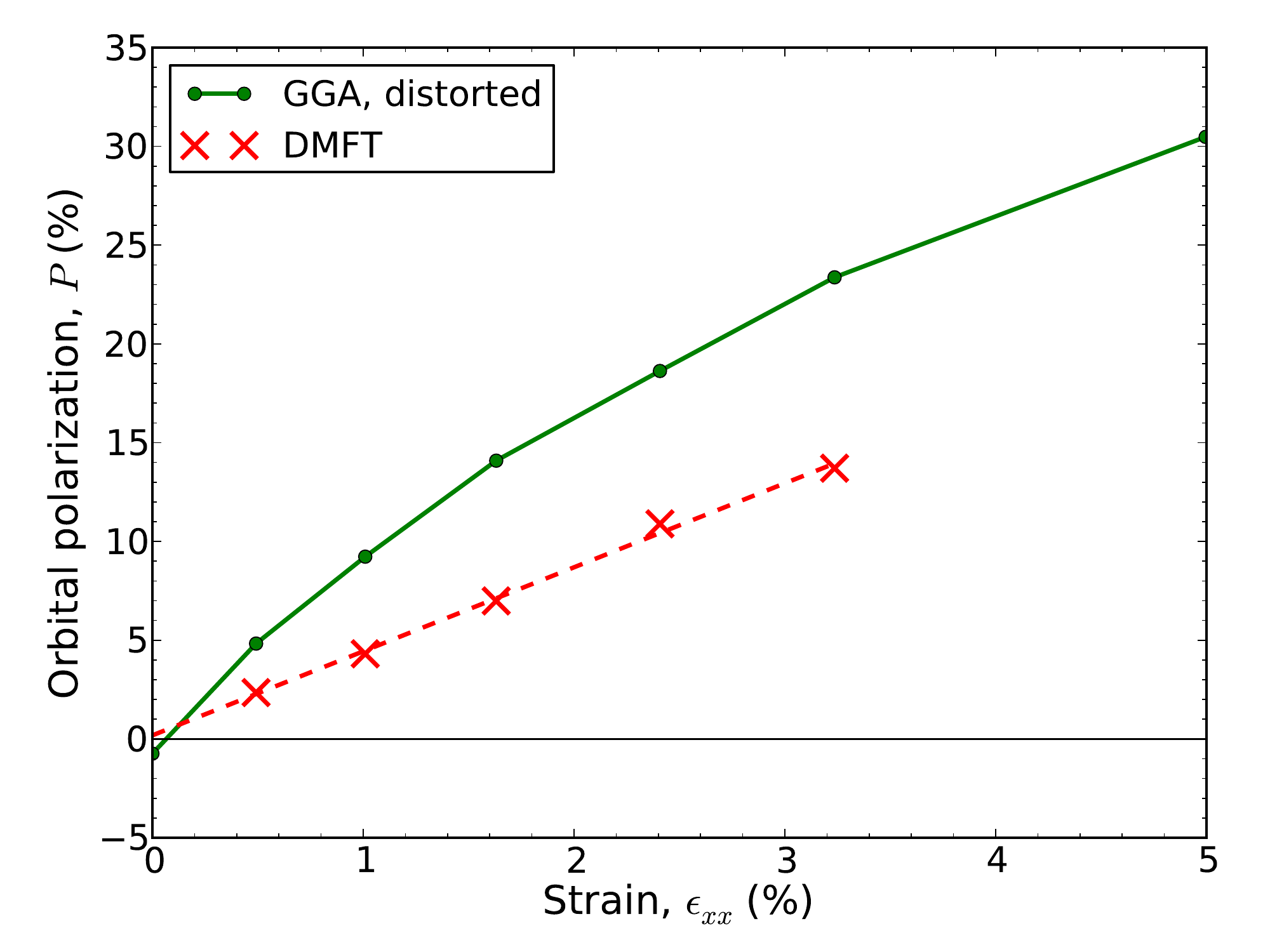}
\caption{(Color online) OP as a function of strain obtained within GGA (green solid line)
and within GGA+DMFT (red crosses; dashed line is a linear fit) for
the distorted structure.
}
\label{fig:op_strain}
\end{figure}
%

We have seen that strain-induced structural distortions 
lead to rather large values of the orbital polarization in GGA electronic structure calculations. 
However, as emphasized in the introduction, nickelates are materials with strong electronic correlations. 
In this section, we investigate how these correlations affect the OP and modify the 
band-structure values above.
Comparison to recent experiments will be made in Sec.~\ref{sec:comp}.
%

\subsection{DFT+DMFT results}
\label{sec:dft_dmft}
We have performed calculations for the relaxed (distorted) structure determined
above with a combination of density-functional theory (DFT-GGA) and dynamical
mean-field theory (DMFT)~\cite{Aichhorn2009,Aichhorn2011} using the
Wien2TRIQS~\cite{TRIQS} interface.  The DMFT quantum impurity problem has been
solved with the numerically exact hybridization-expansion continuous-time
quantum Monte Carlo (CT-QMC) method~\cite{Gull2011} implemented in the
TRIQS~\cite{TRIQS} package.
Importantly, localized Wannier (LW) functions are used in defining the many-body Hamiltonian 
and the corresponding DMFT local impurity problem. These LW functions are defined from 
Ni $d$ and O $p$ states within a large energy range $[-8.0, 4.0]$ eV. 
Our GGA+DMFT calculations hence include all relevant oxygen and nickel states, which is 
physically important in view of the strong hybridization between these states (in contrast 
to a calculation starting from a low-energy Hamiltonian constructed from an EW basis).  
Interactions are applied to these local orbitals using a Slater-type parametrization
with on-site Coulomb interaction and Hund's coupling $U=8.0$~eV, $J_H=1.0$~eV, respectively. 
The double-counting term is chosen to be of the around-mean-field (AMF) form, in view of 
the metallic nature of LNO. 
The calculations were performed only for tensile strains stabilizing the 
$x^2-y^2$ orbital (i.e., positive OP, as defined above). 

As emphasized above however, the $\eg$ occupancies entering expression (\ref{eq:op}) must 
be defined with respect to a basis of extended Wannier functions (EW) as defined in 
Sec.~\ref{sec:basis}. In order to comply with this physical requirement
and to compare in a consistent manner the value of the OP obtained 
in GGA+DMFT with the band structure GGA results, as well as with experiments, 
the local Green's function obtained in the GGA+DMFT calculation in the LW basis is thus reprojected 
onto the EW basis. Details of this procedure are provided in Appendix A.

The resulting orbital polarization as a function of strain is presented in
Fig.~\ref{fig:op_strain} along with corresponding GGA values.
For all values of tensile strain we observe that correlations tend to 
reduce the OP as compared to the GGA values. This finding is in agreement 
with previous calculations of LNO-based heterostructures using DMFT with a similar
choice of the localized Wannier basis,\cite{Han2011} 
although quantitative comparison is difficult in view of the difference
in the systems studied.
As noted in the introduction, DMFT calculations performed using the very different framework of 
a low-energy description involving only $\eg$ states (defined, e.g., from an EW basis) 
found a large enhancement of the OP by correlation effects.\cite{Hansmann2009,Hansmann2010} 
This raises the important question of what is the appropriate minimal
low-energy model for nickelates,\cite{Han2010}
a question that is beyond the scope of the present paper but
which we intend to return to in future work.

%
\begin{figure}
\includegraphics[width=1.0\linewidth]{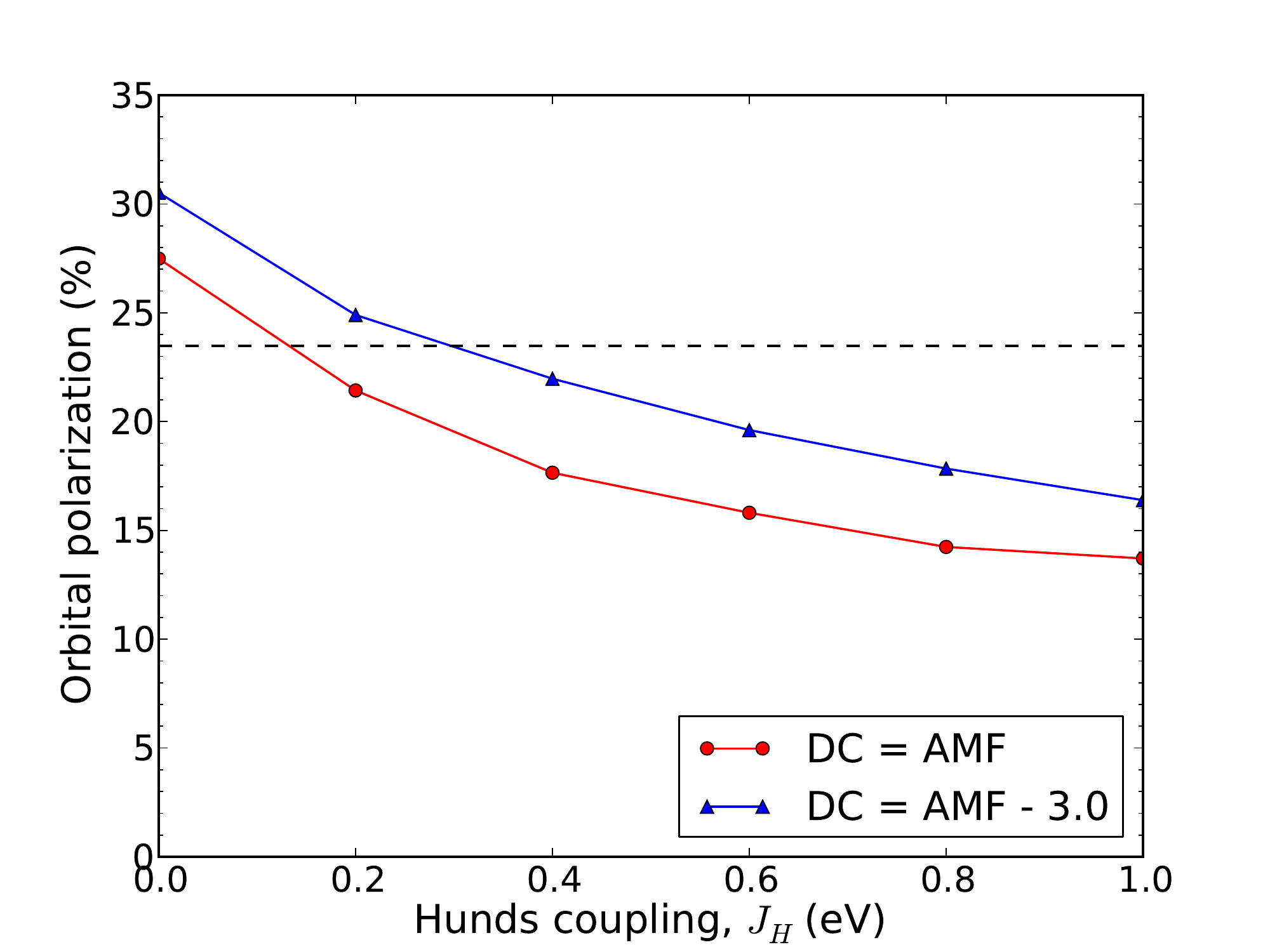}
\caption{Orbital polarization as a function of Hund's rule 
coupling $J_{H}$ for $U=8.0$~eV, obtained within GGA+DMFT for LNO with tensile strain
$\epsilon_{xx} = 3.24\%$. 
The dashed horizontal line indicates the GGA value.  
The results are displayed for two values of the double counting: the 
around mean-field (AMF) one and one in which the double-counting potential is 
shifted by $-3$~eV from the AMF value 
(the shift of the double counting can be viewed as a static contribution of $U_{pd}$ within
Hartree-Fock approximation).
}
\label{fig:op_hunds}
\end{figure}
%

\subsection{Hund's rule coupling and the reduction of orbital polarization}
We now provide a physical discussion of the correlation-induced mechanisms that 
tend to suppress orbital polarization. 

As emphasized in model studies (see, e.g., Ref.~\onlinecite{Poteryaev2008}), 
the on-site repulsion $U$ has a tendency to {\it increase} the OP. This is because 
the crystal-field splitting should be compared to the electronic kinetic energy, which is 
reduced when $U$ (and hence the quasiparticle bandwidth) is increased. Hence, the 
orbital polarizability, proportional to the inverse of the kinetic energy, 
is increased by this effect. 

In contrast, the Hund's rule coupling competes with the crystal-field splitting and 
tends to reduce the OP, as also emphasized in previous studies (for a review, see, 
e.g., Ref.~\onlinecite{Georges2013}). 
These opposite effects of $U$ and $J_H$ are clearly illustrated by Fig.~\ref{fig:op_hunds}, in 
which we display the results of GGA+DMFT calculations performed at several values of 
the Hund's coupling, for LNO subject to the largest tensile strain considered in this work. 
For small $J_H$, the OP is enhanced by correlations as compared to the GGA value, while 
increasing $J_H$ quickly brings the OP down to values smaller than the GGA ones. 

We now discuss qualitatively the physical origin of this effect of the Hund's coupling, 
starting from the atomic limit. 
The ground state configuration of Ni ions can be described as a mixture
of $d^{7}$ and $\dlbar$ configurations (see Sec.~\ref{sec:basis}).
\cite{Abbate2002,Liu2011,Mizokawa1994}
In a $\dlbar$ configuration with two $d$ electrons in the two $\eg$ orbitals, 
the OP will be suppressed because the CF splitting has to compete with the 
rather large Hund's coupling $J_{H} \sim 1$~eV which tends to put 
the two electrons in different orbitals, in a high-spin $S=1$ state.  
The orbital polarizability is obviously zero in such a state.
The relevance of the $\dlbar$ configuration in explaining the observed reduction of the OP 
by the Hund's rule coupling has been emphasized in Ref.~\onlinecite{Han2011}. 

Nevertheless, this should not be taken as evidence that the $\dlbar$ configuration dominates the 
wave-function of Ni ions. Indeed, as we now explain, the Hund's rule coupling also 
acts to reduce the OP for the $d^7$  component of the wavefunction. 
At first sight, this statement appears surprising: obviously, the competition of Hund's rule coupling 
and CF splitting is absent in the atomic limit of individual atoms, since the $d^7$ configuration corresponds 
to a completely filled $t_{2g}$ shell and one electron in the doubly degenerate $e_g$ shell. For an isolated 
atom in this configuration, a small crystal field can fully polarize the orbital configuration 
irrespective of $J_H$. 

\begin{figure}
\includegraphics[width=1.0\linewidth]{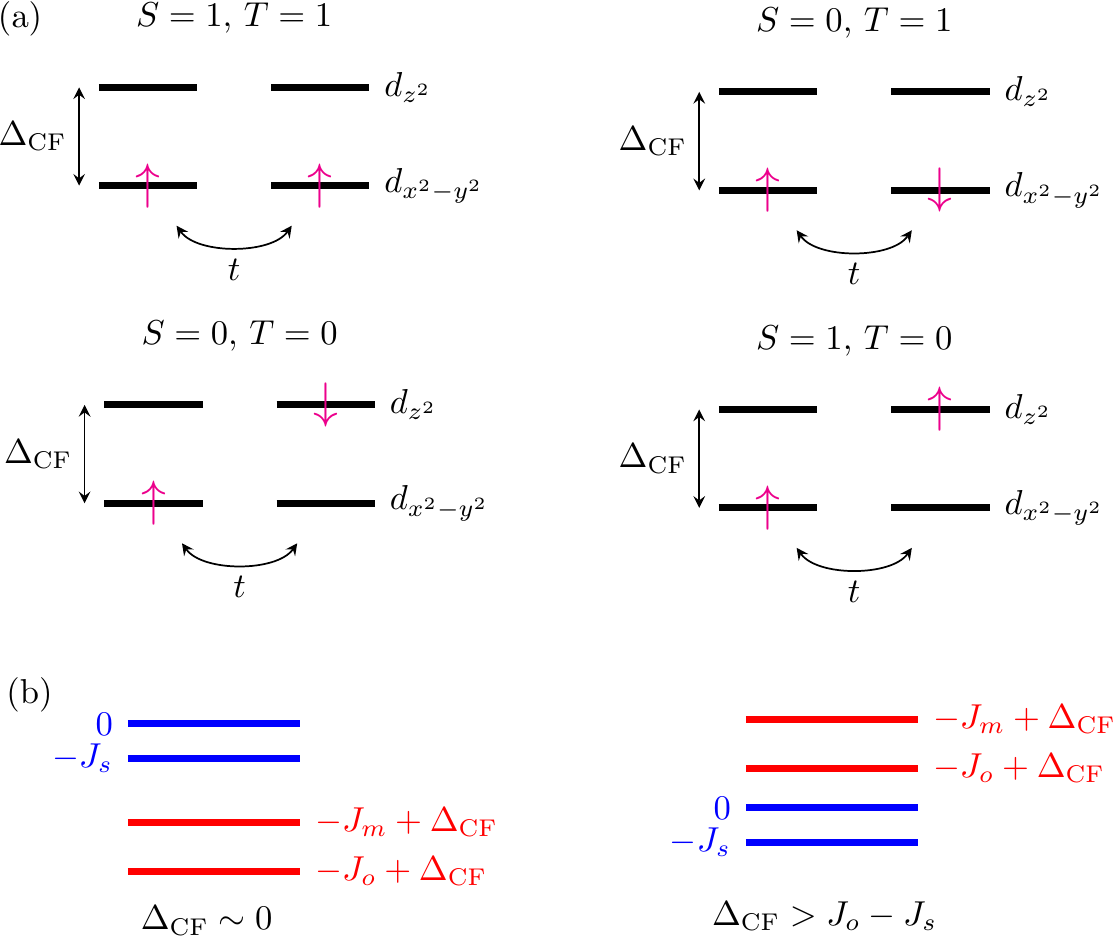}
\caption{(Color online) (a): Four types of configurations of the two-site model corresponding to
four combinations of quantum numbers $S$ and $T$.
(b): Many-body low-energy spectrum produced by the superexchange processes
for small CF splitting (left) and for large CF splitting (right).
}
\label{fig:exch}
\end{figure}

However, this no longer applies when inter-site hopping is taken into account. 
In order to illustrate this point, we consider a simple two-site model (Fig.~\ref{fig:exch}). 
Each site carries two orbitals (i.e., we do not consider the filled $t_{2g}$ orbitals), 
a hopping $t$ connects only orbitals of the same type from one site to another, and 
there is one electron per site on average ($d^7$ configuration, corresponding to one 
electron in the $e_g$ shell).  The levels on each site are split by a CF $\cf$.  
There are $4^2=16$ states in the low-energy Hilbert space with no double occupancy. 
A given eigenstate in this low-energy subspace is characterized by the total spin $S=0,1$ and 
the total orbital pseudo-spin $T=0,1$ (with $T=0$ corresponding to an orbitally 
degenerate state and $T=1$ to an orbitally polarized one). 

In the limit $t \ll U$ the low-energy dynamics is determined
by three types of superexchange processes shown in Fig.~\ref{fig:exch}(a),
with the parameters
\begin{align*}
(S = 1,\, T = 1):\, & \textrm{no hopping is allowed}, \\
\textrm{spin }(S=0,\, T=1):\, & J_{s} = \frac{4t^{2}}{U}, \\
\textrm{orbital }(S=1,\, T=0):\, & J_{o} = \frac{4t^{2}}{U - 3J_{H}}, \\
\textrm{mixed }(S=0,\, T=0):\, & J_{m} = \frac{4t^{2}}{U - 2J_{H}}.
\end{align*}

The superexchange splits the original 16 configurations into four groups of
degenerate states corresponding to four possible combinations of total spin
and pseudo-spin moments [see Fig.~\ref{fig:exch}(b)].
Choosing the energy of the $(S=1,\, T=1)$ configuration as zero and evaluating
the energies of the three other configurations, one readily obtains 
that depending on the value of $\cf$ the ground-state configuration 
will be one of the two following ones: 
\begin{align*}
S = 0,\, T = 1\;\textrm{with energy}\; & E = -J_{s},\\
S = 1,\, T = 0\;\textrm{with energy}\; & E = \cf - J_{o},
\end{align*}
Hence, we conclude that for small CF $\cf < J_{o} - J_{s}$ 
(i.e. for $\cf < 12 t^2 J_H/U^2$ to first order in $J_H/U$), 
the orbitally degenerate configuration $(S=1,\, T=0)$ is the ground state, 
while orbital polarization takes over above this critical value. 
Hence, increasing $J_H$ does increase the stability of the orbitally degenerate 
(unpolarized) state due to the effect of inter-site spin and orbital superexchange.
As a side remark, we mention that even though DMFT considers a single-site effective 
problem, it does capture these inter-site effects in the response of the system to a {\it uniform} field 
(coupling either to orbitals or spin degrees of freedom), as explained in Ref.~\onlinecite{Poteryaev2008}. 

This analysis of course applies to the strong-coupling localized limit of small hopping, and
should not be applied quantitatively to LNO, which is a metal. 
However, it does make the point that Hund's coupling acts to reduce the 
OP even when the nominal atomic configuration is $d^7$. 
At weak coupling, a perturbative analysis leads to similar qualitative conclusions: 
the orbital polarizability is enhanced by correlations, as compared to the free-electron one,   
when $J_H<U/5$, while it is suppressed for $J_H>U/5$ 
(see Ref.~\onlinecite{Georges2013}).  

We conclude that the Hund's coupling-induced suppression of the orbital polarization 
actually applies to both the $\dlbar$ and $d^7$ configurations.
In the latter case, the suppression is mediated by intersite fluctuations involving
virtual $d^{8}$ states.
This suppression should 
not thus be taken as evidence that $\dlbar$  is the dominant component of 
the wave-function in LNO (although it certainly has a sizeable weight, 
in view of the strong hybridization with oxygen states). 
Finally, we note that the analysis of the simple two-site model above shows that 
the Hund's coupling tends to promote an inter-site ferromagnetic alignment 
of spins in the orbitally compensated state with $T=0$, due to the 
inter-site orbital superexchange. 
Indeed, the magnetic susceptibility of LNO displays a large 
Stoner enhancement in \lanio.\cite{Sreedhar1992,Xu1993,Zhou2000} 
%

%
%

\section{Comparison to spectroscopy experiments}
\label{sec:comp}
In this section, we compare our DFT+DMFT results for the orbital polarization as a 
function of strain to the recent experimental results of Wu {\it et al.}\cite{Wu2013}
In these experiments, both dichroism measurements in XAS and  
resonant reflectometry were performed on (LNO)$_{4}$/(RMO)$_{4}$ heterostructures, 
for various substrates RMO$_{3}$ (corresponding to different imposed strains).
With the aid of sum rules the ratio of hole occupancies of $\eg$-states,
$X = h_{\zz} / h_{\xxyy}$, was extracted. 
This ratio can be converted into the OP as defined in Eq.~\eqref{eq:op} by the relation
\begin{equation}
P = \left(\frac{4}{n_{\eg}} - 1\right) \frac{X - 1}{X + 1},
\end{equation}
This expression involves the total occupancy in the $\eg$ states, a quantity that is not 
directly accessible to these experiments. In Ref.~\onlinecite{Wu2013} a value
$n_{\eg} = 1$
corresponding to low-energy Wannier construction (EW basis)
was used with a goal to compare the values of $P$ with the theoretical values
obtained by integrating the DOS of low-energy anti-bonding states.\cite{Han2011}
The XAS measurements yield a value of the OP, $\Pav$, averaged over four layers
of LNO in the heterostructure. To disentangle interfacial and strain effects,
additional measurements using $q$-resolved resonant reflectivity were performed,  
which allowed the experimentalists to obtain the OP for individual layers: two $B$ layers
adjacent to the interface with RMO and two inner $A$ layers in between the $B$ layers,
with corresponding OPs $P_{A}$ and $P_{B}$ that average to $\Pav$.

\begin{figure}
\includegraphics[width=1.0\linewidth]{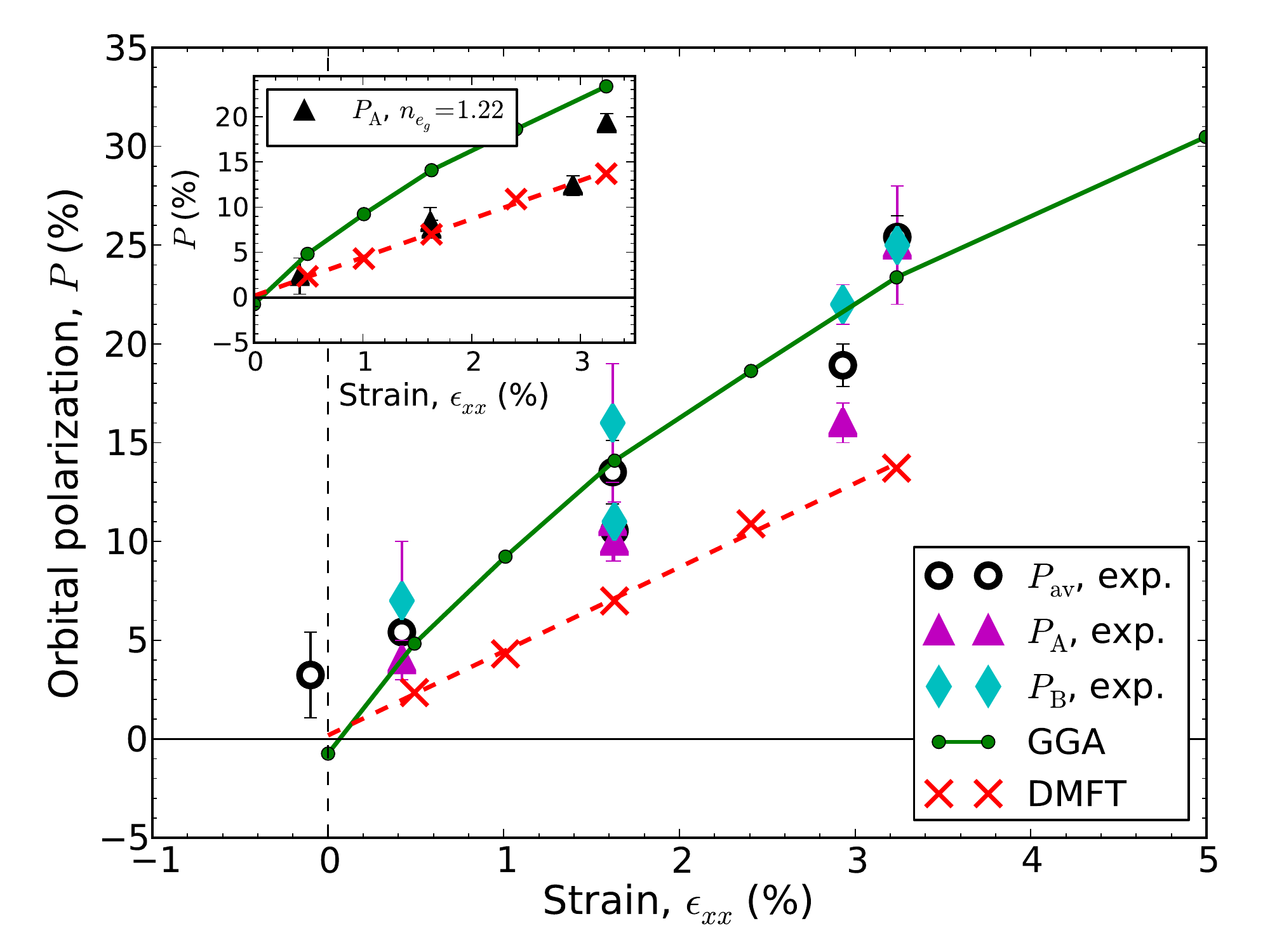}
\caption{(Color online) The orbital polarization, $P$, as a function of strain for the
distorted structure (circles, green), GGA+DMFT (red crosses), and
experiment, with
diamonds, triangles and open circles corresponding respectively to $P_{\textrm{B}}$,
$P_{\mathrm{A}}$, and $P_{\mathrm{av}}$ obtained from XLD in Ref.~\onlinecite{Wu2013}.
Inset: Reinterpretation of the experimental data with a theoretical
value of $n_{\eg} = 1.22$.}
\label{fig:op_exp}
\end{figure}

The experimental results along with the calculated OP already presented in the
previous section are shown together in Fig.~\ref{fig:op_exp}.
As the geometry employed in our calculations takes into account only
a uniform biaxial strain and assumes no interfacial effects, the numerical results
should be compared to the $A$ layer data ($P_{A}$), which are less influenced by the interface.
With this in mind, one can see that pure GGA values substantially
overestimate the polarization and that the agreement between our GGA+DMFT results 
and experimental values is fairly good.

\begin{figure}
\includegraphics[width=1.0\linewidth]{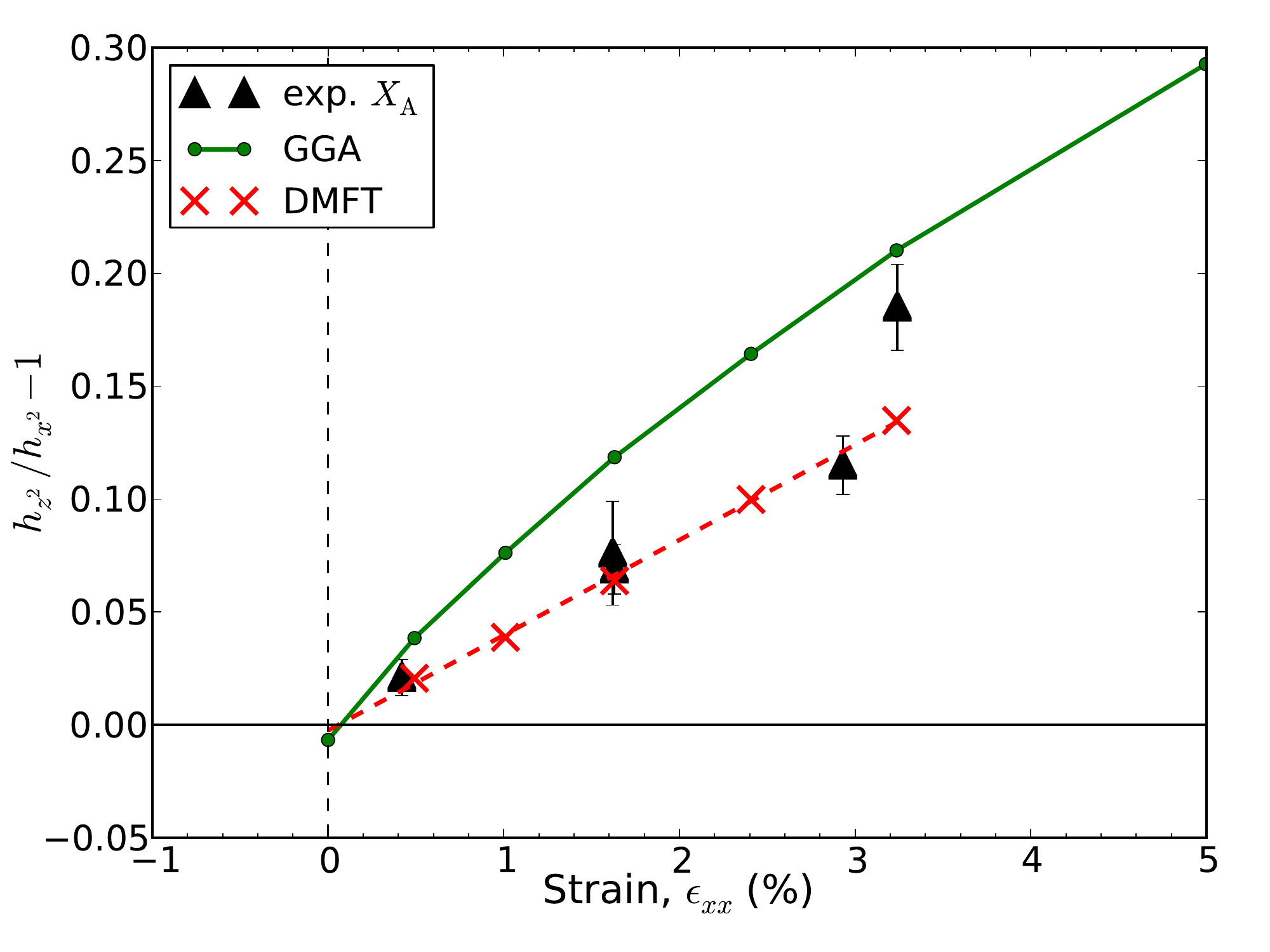}
\caption{(Color online) The ratio of the hole occupancies, $X$,
as a function of strain (for experimental points the value for the inner
layers, $X_{A}$, is used).}
\label{fig:rhole}
\end{figure}

This comparison between our theoretical results and experiments can actually be further refined 
by noting that 
the interpretation of both the hole $h_{\eg}$ and the total $\eg$ occupancies
is basis dependent, as has already been discussed in Sec.~\ref{sec:basis}.
The ambiguity about $n_{\eg}$ can be resolved by simply using
the value of $X$, as originally introduced in Ref.~\onlinecite{Wu2013},
rather than $P$ when comparing calculations with experiment. 
This is done in Fig.~\ref{fig:rhole} where we show theoretical values of $X$ and 
experimental layer-resolved values, $X_{A}$.
Interestingly, there is a noticeable improvement in the agreement between our theoretical 
results and experimental values when the comparison is done in this manner.

The difference between these two analyses of $P$ and $X$ 
can be traced back to the deviation of the actual $\eg$ occupancy
as calculated in GGA+DMFT from the nominal value $n_{\eg}=1$ assumed in 
the analysis of the experimental data in Ref.~\onlinecite{Wu2013}. 
Indeed, if we take the GGA+DMFT value $n_{\eg} \approx 1.22$ (practically independent of strain)
and reinterpret the experiments by recalculating the electron OP from the measured value 
of $X_{A}$ (inset of Fig.~\ref{fig:op_exp}), we get a noticeable shift 
of the data and a correspondingly improved agreement for $P$. 
This reflects the improved agreement obtained when comparing directly the measured 
hole-occupancy ratio $X$.

To summarize, the approach of comparing hole-occupancy ratio, $X$
[or, equivalently, hole orbital polarization $(X - 1)/(X + 1)$] directly
with experiment has a two-fold advantage.
On one hand, one avoids relying on the value of $n_{\eg}$ -- 
a quantity poorly defined from the experimental point of view.
On the other hand,
the energy scale of hole occupancies of $d$  or mixed $d$-$p$ states
is uniquely fixed by the extent of unoccupied
(anti-bonding) states above the Fermi level, which makes
the hole OP independent of the choice of the integration limits
taken when evaluating the occupancies. 

Also, we note that the hole-occupancy ratio $X$ is less sensitive to
the choice of projectors (localized- or extended-Wannier) used in the
evaluation of occupancies.
Within DFT, the low-energy O$p_{\sigma}$ states have the same symmetry
and almost the same positions
as corresponding Ni $\eg$ states, which makes their non-interacting
DOS very similar in shape. The ratio of hole occupancies
evaluated within DFT is thus practically independent of the
type of projectors used. In DMFT calculations within LW basis,
correlations mostly affect $d$ states, shifting their positions and
renormalizing the CF splitting. However, even in this case
the difference between $X_{\sw}$ and $X_{\bw}$ does not exceed
50\% of $X_{\sw}$.

%
%
%
%

%
%
\section{Conclusion}

We have investigated the effect of structural distortions and relaxation, as 
well as the effect of electronic correlations on the orbital polarization 
of \lanio epitaxial films. 

From the structural point of view, we have emphasized the interplay between distortions and 
relaxation in this material. Under tensile strain, in-plane rotations are suppressed and 
out of plane tilts are favored.\cite{May2010} Surprisingly, this leads to a 
larger orbital polarization than for a tetragonally constrained structure subject 
to the same strain, even though the intra-$e_g$ crystal-field splitting is comparatively 
smaller for the relaxed distorted structure. We have shown that this effect is due 
to the mixing of $\eg$ states with low-dispersion $\ttg$ states in the distorted structure. 

This effect by itself would lead to rather large values of the OP, larger than those 
reported in the recent experiment of Wu \textit{et al.}\cite{Wu2013}
Electron correlations lead to a reduction of the OP however, due to the 
effect of the Hund's rule coupling, as previously emphasized by Han \textit{et al.}\cite{Han2011}
This suppression is often interpreted as a signature of the dominance of the
$\dlbar$ configuration in the ground state: in this configuration the CF splitting has
to overcome the Hund's exchange $J_{H}$ to induce an OP and a concomitant high-spin to 
low-spin transition. 
However, we point out that the Hund's exchange also competes with the CF for 
the $d^{7}$ configuration because the strength of the inter-site orbital
superexchange depends on the Hund's coupling.
 
Our theoretical results for the orbital polarization as a function of strain
are in good agreement with the experimental values reported by Wu \textit{et al.}\cite{Wu2013}
We have also emphasized that  a more direct comparison to linear dichroism XAS experiments  
(and better agreement with the experimental data) is achieved when directly comparing 
the ratio of hole occupancies. 

The presented results suggest that although the OP in \lanio can be efficiently
controlled by crystal-structure design, 
achieving a higher degree of orbital polarization is hampered by the three following effects.
\renewcommand{\theenumi}{(\roman{enumi})}
\begin{enumerate}
\item Degeneracy of the two $\eg$ bands at the $\Gamma$-point in \lno strained film geometry
due to vanishing $d$-$p$ hybridization and practically isotropic direct hoppings
in plane and out of plane typical of an ABO$_{3}$ perovskite structure.
This problem can be circumvented 
either by choosing a less symmetric bulk structure (such as an A$_{2}$BO$_{4}$-type structure) or
by engineering
heterostructures with a single
layer of LNO sandwiched between insulating layers;\cite{Chaloupka2008,Yang2010,Han2010,Chen2013b}
however, such structures are generally difficult to fabricate.
\item Small charge-transfer energy (possible even negative\cite{Mizokawa2000}),
resulting in a significant contribution of
the $\dlbar$ configuration.
\item The reduction of the OP by the Hund's rule coupling.
\end{enumerate}
Further progress in the field aiming at achieving larger or even full
orbital polarization\cite{Chaloupka2008,Hansmann2009,Hansmann2010} 
will have to overcome these effects by considering appropriate materials and heterostructures. 

\begin{acknowledgments}
We are grateful to Sara Catalano, Stefano Gariglio, Marta Gibert, Jean-Marc Triscone and 
all other members of the Triscone group in Geneva, as well as to Philipp Hansmann, Andrew J. Millis, 
Leonid Pourovskii, Dirk van der Marel, George Sawatzky and Michel Viret, for many useful discussions.
We are indebted to Eva Benckiser, Meng Wu and Bernhard Keimer for discussions 
about their data and for making them available to us.   
Support for this work was provided by the Swiss National Science Foundation (Grant No. 20021-146586)
and MaNEP, by a grant from the European Research Council (ERC-319286 QMAC),
and by the Swiss National Supercomputing Centre (CSCS) under Project ID s404. 
\end{acknowledgments}

%
%
\section*{Appendix A}
Here we outline the routine of reprojecting the impurity Green's function (GF)
to get occupation numbers from DMFT calculations that are consistent
with those obtained previously in pure GGA calculations.
The local impurity problem in the DMFT self-consistency cycle is constructed using
projectors $P^{\bw}_{m\nu}(\kv)$ defined within energy window $[-8.0, 4.0]$ eV
(as described in Sec.~\ref{sec:dft_dmft}),
which defines the resulting (converged) impurity GF as
\begin{widetext}
\begin{align*}
\GlocBW_{mm'}(\iomn) = & \sum_{\kv\nu\nu'} P^{\bw}_{m\nu}(\kv) 
\Gband_{\nu\nu'}(\kv, \iomn) \left[P^{\bw}\right]^{*}_{\nu'm'}(\kv), \\
\Gband_{\nu\nu'}(\kv, \iomn) = & \left(
[\iomn + \mu - \eps_{\kv\nu}]\delta_{\nu,\nu'} - \sum_{mm'}
\left[P^{\bw}\right]^{*}_{\nu m}(\kv) \Sigloc_{m m'}(\iomn) P^{\bw}_{m'\nu'}(\kv)
\right)^{-1},
\end{align*}
\end{widetext}
where $\Sigloc_{m m'}(\iomn)$ is the converged self-energy and indices
$m$ run over all five $d$ orbitals.

The occupation numbers and the OP obtained directly from this impurity GF
will be inconsistent with the OP given in Sec.~\ref{sec:op_gga} as 
$\Sigloc_{m m'} \to 0$ because of the difference in the basis sets.
To get consistent occupation numbers, $n_{l}$, with index $l = 1,2$ running
only over $\eg$-orbitals, we project the above band GF,
$\Gband_{\nu\nu'}(\kv, \iomn)$ using the EW basis (energy window
$[-1.6, 4.0]$ eV) projectors,
$P^{\sw}_{l\nu}(\kv)$,
\begin{widetext}
\begin{align*}
n_{l} = & \frac{1}{\beta} \sum_{\iomn} \GlocSW_{l l}(\iomn)e^{\iomn 0^{+}}, \\
\GlocSW_{l l'}(\iomn) = & \sum_{\kv\nu\nu'} P^{\sw}_{l\nu}(\kv) 
\Gband_{\nu\nu'}(\kv, \iomn) \left[P^{\sw}\right]^{*}_{\nu'l'}(\kv).
\end{align*}
\end{widetext}

%
%
\section*{Appendix B}
%
\begin{figure}
\includegraphics[width=1.0\linewidth]{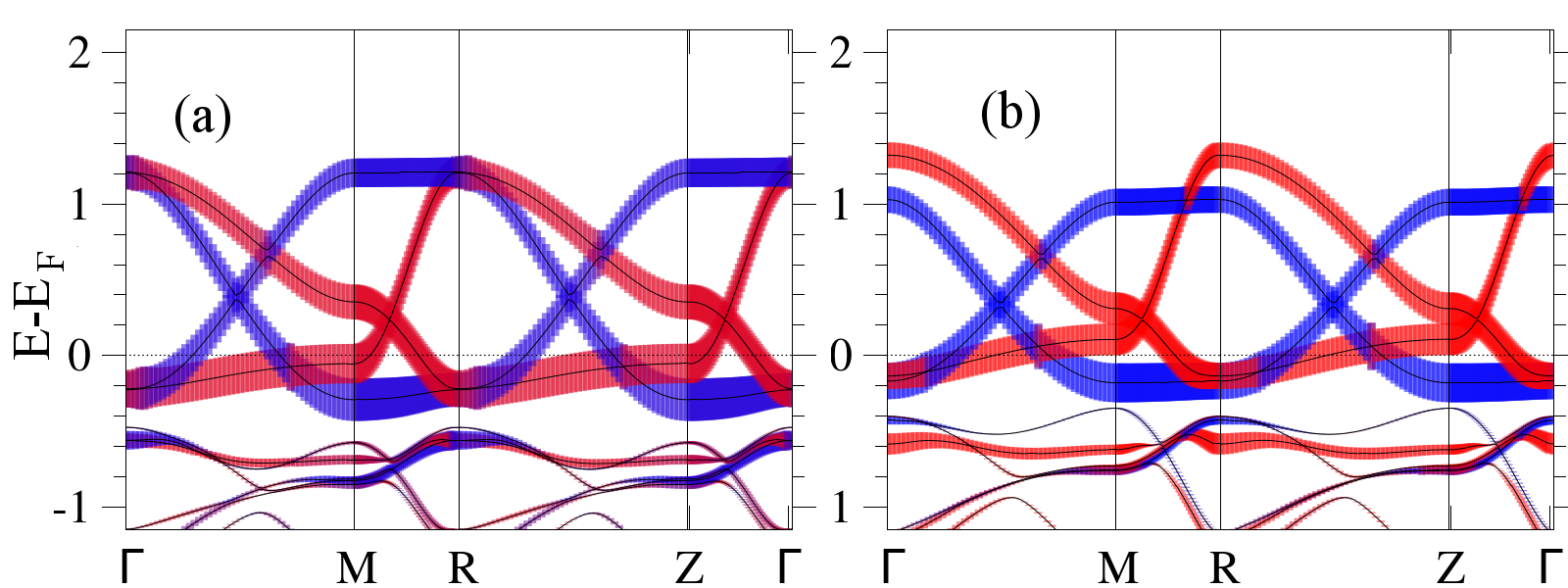}
\caption{(Color online) Renormalized band structure of quasi-particles obtained using GGA+DMFT
(distorted structure): (a) $0.0\%$ strain (b) $+3.2\%$ strain. 
The figures are to be compared with 
Figs.~\ref{fig:bs_unstrained},~\ref{fig:bs_strained} (note that the
scale of the ordinate is different here).}
\label{fig:qp_bands}
\end{figure}
%

To demonstrate the effect of correlations on the band structure we present
the quasi-particle band structure resulting from GGA+DMFT calculations
for two cases: unstrained [Fig.~\ref{fig:qp_bands}(a)] and for tensile strain [Fig.~\ref{fig:qp_bands}(b)].
Apart from an overall narrowing of the bands in both cases, one can see a significant lifting of the 
$\eg$ states at the $\Gamma$ point. There is also a change in the Fermi surface topology
visible along the $\Gamma-M$ line, which is in accord with recent results from
angular-resolved photoemission spectroscopy.\cite{Yoo2014} In the experimental work,
the appearance of the hole pocket at the $M$ point was attributed to correlation
effects in a sample under tensile strain. The pockets, however, appear already in
GGA calculations if one considers a fully relaxed distorted structure, as can
be seen from comparing the band structures for tetragonal and distorted cases
in Figs.~\ref{fig:bs_unstrained} and \ref{fig:bs_strained}.

%

\end{document}